\documentclass{jfm}
\usepackage{graphicx}
\usepackage{newtxtext}
\usepackage{newtxmath}
\usepackage{natbib}
\usepackage{hyperref}
\usepackage{xcolor}
\usepackage{pifont}
\usepackage{tikz}
\usetikzlibrary{tikzmark}
\tikzset{mycircled/.style={circle,draw,inner sep=0.1em,line width=0.1em}}
\hypersetup{
    colorlinks = true,
    urlcolor   = blue,
    citecolor  = blue,
}

\newcommand{\RomanNumeralCaps}[1]
\linenumbers

\shorttitle{Viscoelastic fluid flow in a contraction: the role of finite extensibility}
\shortauthor{B. Mahapatra, T. Ruangkriengsin, H.A. Stone and E. Boyko}

\title{\Large Viscoelastic fluid flow in a slowly varying planar contraction: the role of finite extensibility on the pressure drop}
 \author{ Bimalendu Mahapatra\aff{1},
Tachin Ruangkriengsin\aff{2},
 Howard A. Stone\aff{3} \and 
 Evgeniy Boyko\aff{1}
 \corresp{\email\href{mailto:evgboyko@technion.ac.il}{evgboyko@technion.ac.il}}
}
  \affiliation{\aff{1}Faculty of Mechanical Engineering, Technion – Israel Institute of Technology,
Haifa 3200003, Israel
\aff{2}Program in Applied and Computational Mathematics, Princeton University, Princeton, NJ 08544, USA
\aff{3}Department of Mechanical and Aerospace Engineering, Princeton University, Princeton, NJ 08544, USA}

\begin{document}
\maketitle

\begin{abstract}

We analyze the steady viscoelastic fluid flow in slowly varying contracting channels of arbitrary shape and present a theory based on the lubrication approximation for calculating the flow rate--pressure drop relation at low and high Deborah ($De$) numbers. Unlike most prior theoretical studies leveraging the Oldroyd-B model, we describe the fluid viscoelasticity using a FENE-CR model and examine how the polymer chains' finite extensibility impacts the pressure drop. We employ the low-Deborah-number lubrication analysis to provide analytical expressions for the pressure drop up to $O(De^4)$. We further consider the ultra-dilute limit and exploit a one-way coupling between the parabolic velocity and elastic stresses to calculate the pressure drop of the FENE-CR fluid for arbitrary values of the Deborah number. Such an approach allows us to elucidate elastic stress contributions governing the pressure drop variations and the effect of finite extensibility for all $De$. We validate our theoretical predictions with two-dimensional numerical simulations and find excellent agreement. We show that, at low Deborah numbers, the pressure drop of the FENE-CR fluid monotonically decreases with $De$, similar to the previous results for the Oldroyd-B and FENE-P fluids. However, at high Deborah numbers, in contrast to a linear decrease for the Oldroyd-B fluid, the pressure drop of the FENE-CR fluid exhibits a non-monotonic variation due to finite extensibility, first decreasing and then increasing with $De$. Nevertheless, even at sufficiently high Deborah numbers, the pressure drop of the FENE-CR fluid in the ultra-dilute and lubrication limits is lower than the corresponding Newtonian pressure drop.

\end{abstract}

\begin{keywords}
 non-Newtonian flows, viscoelasticity, lubrication theory
\end{keywords}


\section{Introduction}
The ability to accurately predict the hydrodynamic features is at the core of understanding viscoelastic fluid flows. 
Such complex fluid flows may exhibit significantly different characteristics from Newtonian flows, even with a small concentration of polymer molecules present, giving rise to viscoelastic effects such as normal stress differences and extensional thickening~\citep{bird1987dynamics1,steinberg2021elastic,datta2021perspectives,ewoldt2022designing}. 

One hydrodynamic feature that has received considerable attention in the fluid mechanics community is the relationship between the pressure drop $\Delta p$ and the flow rate $q$ in viscoelastic channel flows with spatially
varying shapes. Over the years, the $q- \Delta p$ relation of viscoelastic fluid flows has been studied in different geometries,  through numerical simulations~\citep{szabo1997start,alves2003benchmark,binding2006contraction,alves2007divergent,zografos2020viscoelastic,varchanis2022reduced} and experimental measurements~\citep{rothstein1999extensional,rothstein2001axisymmetric,sousa2009three,ober2013microfluidic,james2021pressure}, and recently, via theoretical analysis~\citep{perez2019analytical,boyko2022pressure,housiadas2023_2D,housiadas2023_Axi,housiadas2024pressure,BoykoHinchStone2023,HinchBoykoStone2023}. For an overview of recent studies, the reader is referred to~\citet{boyko2022pressure} and~\citet{HinchBoykoStone2023}. 

The majority of previous numerical and experimental studies on the flow rate–pressure drop relation have focused on rapidly varying geometries with sharp corners, such as abrupt or hyperbolic contraction and contraction--expansion (constriction) channels~\citep[see, e.g.,][]{rothstein1999extensional,alves2003benchmark,binding2006contraction,campo2011flow,keshavarz2016micro,zografos2022viscoelastic}. However, such rapidly varying geometries greatly complicate theoretical analysis. Therefore, 
to overcome this issue and enable asymptotic analysis, theoretical studies have considered instead a slowly varying geometry and exploited the narrowness of the geometry through the application of the lubrication theory~\citep[see, e.g.,][]{boyko2022pressure,housiadas2023_2D,housiadas2023_Axi}. 
There have been numerous applications of lubrication theory to other viscoelastic fluid flows, such as thin films and tribology problems~\citep{ro1995viscoelastic,tichy1996non,sawyer1998non,zhang2002surfactant,saprykin2007free,ahmed2021new,gamaniel2021effect,datt2022thin,ahmed2023modeling}, as well as translation of a sphere near a rigid plane~\citep{Ardekani_2007,Ruangkriengsin2024} and analysis of forces and torques acting on nearly touching spheres~\citep{dandekar2021nearly}. 

Using such a theoretical approach in conjunction with applying a
perturbation expansion in powers of the Deborah number $De$ (see definition in $\mathsection$~\ref{Scaling}),  \citet{boyko2022pressure} studied the steady flow of an Oldroyd-B fluid in a slowly varying, arbitrarily shaped 2-D channel and provided the expression for the $q- \Delta p$ relation up to $O(De^3)$ in the low-Deborah-number limit. Recently,~\citet{housiadas2023_2D} extended the analysis of~\citet{boyko2022pressure} to much higher asymptotic orders and provided analytical expressions for the pressure drop up to $O(De^8$) for different constitutive models, such as Oldroyd-B, Phan-Thien$-$Tanner (PTT)~\citep{thien1977new,phan1978nonlinear}, Giesekus~\citep{giesekus1982simple}, and a finitely extensible nonlinear elastic (FENE) model with the Peterlin approximation (FENE-P)~\citep{bird1980polymer,bird1987dynamics1}.
Their low-Deborah-number theoretical predictions for pressure drop using more complex constitutive models are very close to those of the Oldroyd-B model, showing a monotonic decrease in the scaled pressure drop with $De$ for the flow through a hyperbolic contraction~\citep{housiadas2023_2D}. 

Recently,~\citet{HinchBoykoStone2023} and \citet{BoykoHinchStone2023} analyzed the flow of an Oldroyd-B fluid in a slowly varying 2-D channel in the high-$De$ limit using lubrication theory. 
\citet{HinchBoykoStone2023}~studied numerically the flow through a contraction, expansion, and constriction for order-one Deborah numbers, and provided asymptotic solutions at high Deborah numbers. \citet{BoykoHinchStone2023} studied the flow of the Oldroyd-B fluid in a slowly varying contraction considering the ultra-dilute limit, in which there is a one-way coupling between the
Newtonian velocity and polymer stresses~\citep{remmelgas1999computational,moore2012weak,li2019orientation,mokhtari2022birefringent}. Such an approach allows for considerable theoretical progress beyond low $De$, yielding semi-analytical expressions for the conformation tensor and pressure drop for arbitrary values of the Deborah number. For a contraction,~\citet{HinchBoykoStone2023} and ~\citet{BoykoHinchStone2023} showed that the pressure drop of the Oldroyd-B fluid monotonically decreases with $De$, scaling linearly with $De$ at high Deborah numbers, and identified two physical mechanisms responsible for the pressure drop reduction.

Although the Oldroyd-B model is the simplest viscoelastic model that combines viscous and elastic stresses and can be derived from kinetic theory, it has several shortcomings~\citep{beris2021continuum,hinch2021oldroyd,shaqfeh2021oldroyd,sanchez2022understanding,StoneShelleyBoyko2023}. One well-known shortcoming of the Oldroyd-B model is that it allows the polymer chains, represented by elastic dumbbells, to be infinitely extensible~\citep{bird1987dynamics1}. However,
in reality, the polymer chains have a finite length.  
More importantly, theoretical and numerical predictions for the pressure drop reduction of an Oldroyd-B fluid in a contraction~\citep{alves2003benchmark,boyko2022pressure,housiadas2023_2D,BoykoHinchStone2023} are in
contrast with the experiments showing a nonlinear increase in the pressure drop with $De$ for the flow of a Boger fluid through abrupt contraction--expansion and contraction geometries~\citep{rothstein1999extensional,rothstein2001axisymmetric,nigen2002viscoelastic,sousa2009three}. As pointed out by~\citet{alves2003benchmark} and \citet{HinchBoykoStone2023}, this discrepancy might be due to the lack of dissipative effects in the Oldroyd-B model. 

Different models, such as the FENE-CR model introduced by~\citet{chilcott1988creeping} and the FENE-P model, incorporate the feature of finite extensibility through a nonlinear restoring force and include extra dissipation.
Similar to the Oldroyd-B model, the FENE-CR model does not account for the shear-thinning effect and is suitable for describing constant shear-viscosity viscoelastic (Boger) fluids~\citep{james2009boger}. In contrast, the FENE-P model incorporates both the finite extensibility and the shear-thinning effect of viscoelastic fluids.

There are several advantages of studying the FENE-CR model prior to the FENE-P model, particularly at high Deborah numbers.
First, the FENE-CR model allows the study of elastic effects on the pressure drop without the influence of shear thinning in shear viscosity. Second, the FENE-CR model is more convenient for theoretical analysis. For example, in contrast to the conformation tensor components of the fully developed flow of a FENE-CR fluid in a straight channel, which have relatively simple expressions (see Appendix~\ref{App: straight channel}), the corresponding expressions for the FENE-P fluid are more cumbersome~\citep{cruz2005analytical}. 

Nevertheless, it should be noted that at low $De$, more complex constitutive models, such as PTT, Giesekus, FENE-P, and FENE-CR, exhibit behavior similar to Oldroyd-B due to the weak effect of additional microscopic features~\citep{BoykoStone2024Perspective}. Indeed, at low Deborah numbers, the PTT, Giesekus, and FENE-P fluids showed only a slight difference in the pressure drop results compared to the Oldroyd-B fluid~\citep{housiadas2023_2D}. However, at high Deborah numbers, 
additional microscopic features, such as finite extensibility, become apparent and impact the elastic stresses~\citep[see, e.g.,][]{zografos2022viscoelastic}. Therefore, one should anticipate significant differences between the predictions for the pressure drop of the Oldroyd-B and the more complex constitutive models, thus motivating further investigation. 

In this work, we study the pressure-driven flow of the FENE-CR fluid in slowly varying, arbitrarily shaped, planar contracting channels using lubrication theory. In contrast to \citet{housiadas2023_2D}, who considered the flow of a FENE-P fluid through a non-uniform channel at low $De$, in current work we analyze the low-Deborah-number limit and the ultra-dilute limit, with the latter enabling us to explore arbitrary values of Deborah number. We first employ a perturbation expansion in powers of the Deborah number 
to calculate the non-dimensional pressure drop for the FENE-CR fluid up to $O(De^4)$ and then apply the Pad\'{e} approximation~\citep{housiadas2017improved} to improve the convergence of the asymptotic series. 
We find that, at low Deborah numbers, the pressure drop of the FENE-CR fluid monotonically decreases with $De$, similar to the Oldroyd-B and FENE-P fluid predictions. To elucidate the pressure drop behavior at high $De$, we consider the ultra-dilute limit of small polymer concentration and leverage a one-way coupling between the parabolic velocity and polymer stresses to calculate the pressure drop for arbitrary values of the Deborah number. 
Such an approach allows us to study the elastic stress contributions governing the pressure drop variations and the effect of finite extensibility for all $De$.
We show that, at high Deborah numbers, in contrast to a linear pressure drop reduction of the Oldroyd-B fluid, the pressure drop of the FENE-CR fluid exhibits a non-monotonic variation, first decreasing and then increasing with $De$. Nevertheless, in the ultra-dilute limit, the pressure drop of the FENE-CR fluid is lower than the corresponding Newtonian pressure drop even at sufficiently high Deborah numbers. We validate our theoretical predictions with 2-D finite-volume numerical simulations and find excellent agreement. However, as expected, at sufficiently high $De$, our 2-D finite-volume numerical simulations, implementing the log-conformation formulation, suffer from accuracy and convergence difficulties due to the high-Weissenberg-number problem~\citep{owens2002computational,alves2021numerical}. Therefore, we believe that our theoretical results for the FENE-CR fluid in the ultra-dilute limit, valid at high Deborah numbers, are of fundamental importance for validating simulation predictions and advancing our understanding of viscoelastic channel flows.

\section{Problem formulation and governing equations}\label{PF}

\begin{figure}
\centerline{\includegraphics[width=\linewidth]{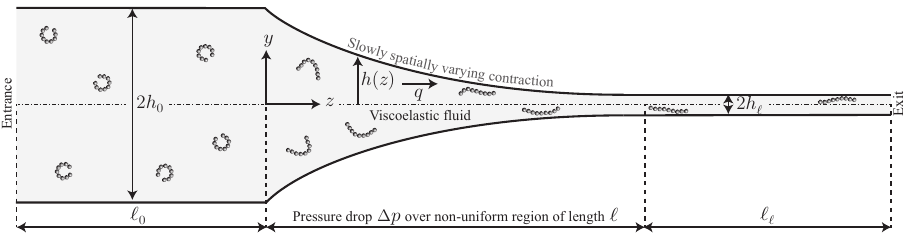}}
\caption{Schematic illustration of the planar configuration consisting of a slowly varying and symmetric contraction of height $2h(z)$ and length $\ell$ ($h\ll\ell$). Upstream of the contraction inlet there is an entry channel of height $2h_{0}$  and length $\ell_{0}$, and downstream of the contraction outlet there is an exit channel of height $2h_{\ell}$ and length $\ell_{\ell}$. 
The flow rate $q$ drives a viscoelastic fluid through the geometry, and we aim to determine the pressure drop $\Delta p$ across the contraction region. 
We have indicated the qualitatively expected extension of polymers as the fluid flows through the contraction since the extension affects the fluid response in the FENE-CR description.}\label{Fig:1}
\end{figure}

We study the incompressible steady flow of a viscoelastic fluid in a slowly varying and symmetric planar channel of height $2h(z)$ and length $\ell$, where $h \ll \ell$, as shown in figure~\ref{Fig:1}. Motivated by the geometries used in previous experimental and numerical studies~\citep[see, e.g.,][]{szabo1997start,rothstein1999extensional,alves2003benchmark,alves2007divergent,campo2011flow,ober2013microfluidic,zografos2020viscoelastic,boyko2022pressure,BoykoHinchStone2023,HinchBoykoStone2023}, 
we assume that the inlet ($z=0$) and outlet ($z=\ell$) of the contraction are connected to two long straight channels of height $2h_{0}$  and $2h_{\ell}$, and length $\ell_{0}$ and $\ell_{\ell}$, respectively. We consider the fluid motion with the pressure distribution $p$ and velocity $\boldsymbol{u} =(u_z, u_y)$ induced by an imposed flow rate $q$ (per unit depth). Our primary interest in this work is to examine the pressure drop $\Delta p$ of a viscoelastic fluid over the contraction region at low and high Deborah numbers while incorporating the finite extensibility of polymer chains.  

We consider low-Reynolds-number flows and neglect the fluid inertia.
In this creeping flow limit, the governing equations are the continuity and
momentum equations
\refstepcounter{equation}
$$
\boldsymbol{\nabla\cdot u}=0,\qquad\boldsymbol{\nabla\cdot\sigma}=\boldsymbol{0}.\eqno{(\theequation{a,b})}\label{Continuity+Momentum FCR}
$$
Here, the stress tensor $\boldsymbol{\sigma}$ can be expressed as
\begin{equation}\label{Stress tensor FCR}
    \boldsymbol{\sigma}=-p\mathsfbi{I}+2\mu_{s}\mathsfbi{E}+\boldsymbol{\tau}_{p},
\end{equation}
where $-p\mathsfbi{I}$ is the pressure contribution, $2\mu_s\mathsfbi{E}$ is the viscous stress contribution of a Newtonian solvent with a constant viscosity $\mu_s$, where $\mathsfbi{E}=(\boldsymbol{\nabla}\boldsymbol{u}+(\boldsymbol{\nabla}\boldsymbol{u})^{\mathrm{T}})/2$ is the rate-of-strain tensor, and  $\boldsymbol{\tau}_p$ is the polymer contribution to the stress tensor. 
 
To describe the viscoelastic rheology of the fluid, we use the FENE-CR model introduced by~\citet{chilcott1988creeping}. In contrast to the Oldroyd-B constitutive equation~\citep{oldroyd1950formulation}, the FENE-CR constitutive model considers polymer molecules as dumbbells
with a finite extensibility $L$ relative to their value
at equilibrium. However, the FENE-CR model does not account for the shear-thinning effect, which can be captured using the FENE-P model~\citep{bird1987dynamics1}.
For the FENE-CR model, the polymer contribution to the stress tensor $\boldsymbol{\tau}_{p}$ can be expressed in terms of the symmetric conformation tensor (or the deformation of the microstructure) $\mathsfbi{A}$ as~\citep{chilcott1988creeping,alves2021numerical},
\begin{equation} \label{tau_p FCR}
    \boldsymbol \tau_p=\frac{\mu_p}{\lambda}F(\mathsfbi{A})(\mathsfbi{A}-\mathsfbi{I}),
\end{equation}
where $\mu_p$ is the polymer contribution to the shear viscosity at zero shear rate and $\lambda$ is the relaxation time. We also introduce the total zero-shear-rate viscosity $\mu_0=\mu_s+\mu_p$.

The function $F(\mathsfbi{A})$ in (\ref{tau_p FCR}) 
accounts for the finite extensibility of polymers represented by elastic
dumbbells and is modeled using the Warner spring function~\citep{warner1972kinetic}, 
\begin{equation}
F(\mathsfbi{A})=\frac{1}{1-(\mathrm{tr}\mathsfbi{A})/L^{2}},\label{Warner spring FCR}
\end{equation}
where $\mathrm{tr}\mathsfbi{A}$ denotes the trace of the conformation tensor $\mathsfbi{A}$.

At a steady state, the conformation tensor  of the FENE-CR model satisfies~\citep{chilcott1988creeping}
\begin{equation}\label{Conformation tensor FCR}
    \boldsymbol{u}\boldsymbol{\cdot}\boldsymbol{\nabla}\mathsfbi{A}-(\boldsymbol{\nabla}\boldsymbol{u})^{\mathrm{T}}\boldsymbol{\cdot}\mathsfbi{A}-\mathsfbi{A}\boldsymbol{\cdot}(\boldsymbol{\nabla}\boldsymbol{u})=-\frac{F(\mathsfbi{A})}{\lambda}(\mathsfbi{A}-\mathsfbi{I}).
\end{equation}For large values of $L$, the function $F(\mathsfbi{A})$ tends to 1, so that the FENE-CR model, given in (\ref{tau_p FCR}) and (\ref{Conformation tensor FCR}), reduces to the steady form of the Oldroyd-B constitutive equation.

\subsection{Non-dimensionalization}\label{Scaling}

We analyze the viscoelastic fluid flow through a
narrow slowly varying channel, in which the channel height is much smaller than the channel length, $h(z)\ll \ell$.
Therefore, for the non-dimensionalization of the viscoelastic flow problem, we introduce dimensionless variables based on the lubrication theory~\citep{tichy1996non,zhang2002surfactant,saprykin2007free,ahmed2021new,ahmed2023modeling,boyko2022pressure,BoykoHinchStone2023}, 
\begin{subequations} \label{Scaling variables FCR}
\begin{equation}\label{Scaling Z, Y, Uz, Uy}
Z=\frac{z}{\ell}, \qquad Y=\frac{y}{h_{\ell}},  \qquad U_z=\frac{u_z}{u_c},  \qquad U_y=\frac{u_y}{\epsilon u_c},
\end{equation}
\begin{equation}\label{Scaling P, delP, Hz}
     \qquad P=\frac{p}{\mu_0 u_c \ell/h^2_{\ell}}, \qquad \Delta P=\frac{\Delta p}{\mu_0 u_c \ell/h^2_{\ell}}, \qquad H(Z)=\frac{h(z)}{h_{\ell}},
\end{equation}
\begin{equation}\label{Scaling A_zz, A_yz, A_yy}
{\tilde A}_{zz}=\epsilon^2 A_{zz}, \qquad {\tilde A}_{yz}=\epsilon A_{yz}, \qquad {\tilde A}_{yy}=A_{yy},
\end{equation}
\begin{equation}\label{Scaling tau_zz, tau_yz, tau_yy}
\mathcal{T}_{p,zz}=\frac{\epsilon^2\ell}{\mu_0 u_c}\tau_{p,zz}, \qquad \mathcal{T}_{p,yz}=\frac{\epsilon \ell}{\mu_0 u_c}\tau_{p,yz},\qquad \mathcal{T}_{p,yy}=\frac{\ell}{\mu_0 u_c}\tau_{p,yy},
\end{equation}
\end{subequations}
where $u_c=q/2h_{\ell}$ is the characteristic velocity scale, $q$ is the imposed flow rate per unit depth, and $h_{\ell}$ is the half-height at $z=\ell$. In addition, we introduce the aspect ratio of the configuration, which is assumed to be
small,
\begin{equation}\label{Aspect ratio FCR}
    \epsilon=\frac{h_{\ell}}{\ell}\ll 1,
\end{equation}
the contraction ratio,
\begin{equation}
H_{0}=\frac{h_{0}}{h_{\ell}},\label{Contraction ratio}
\end{equation}
the viscosity ratios, 
\begin{equation}\label{Viscosity ratio FCR}
    \tilde \beta=\frac{\mu_p}{\mu_s+\mu_p}=\frac{\mu_p}{\mu_0} \quad  \text{and} \quad  \beta=1-\tilde \beta=\frac{\mu_s}{\mu_0},
\end{equation}
and the Deborah and Weissenberg numbers, 
\begin{equation}\label{De_Wi FCR}
    De=\frac{\lambda u_c}{\ell} \quad \text{and} \quad Wi=\frac{\lambda u_c}{h_{\ell}}.
\end{equation}
 Finally, we note that the fluid inertia is negligible, provided the reduced Reynolds number is small,
\begin{equation}\label{Reduced_Reynolds_number FCR}
    \epsilon Re=\epsilon \frac{\rho u_c h_{\ell}}{\mu_0}=\frac{\rho q h_{\ell}}{2\mu_0\ell}\ll1,
\end{equation}
where $\rho$ is the density of the fluid. 

Note that we have defined both the Deborah and Weissenberg numbers. Although the Deborah and Weissenberg numbers are equivalent in many steady flows, in lubrication flows, they have different orders of magnitude due to the two distinct length scales. The Deborah number $De$ is the ratio of the relaxation time of the fluid, $\lambda$, to the residence time in the non-uniform region, $\ell/u_c$~\citep{tichy1996non,zhang2002surfactant,saprykin2007free,ahmed2021new,boyko2022pressure,ahmed2023modeling,housiadas2023_2D,BoykoHinchStone2023,HinchBoykoStone2023}. The Weissenberg number $Wi$ is the product of the relaxation time of the fluid, $\lambda$, and the characteristic shear rate of the flow,  $u_c/h_{\ell}$, and is related to the Deborah number through $De=\epsilon Wi$. Therefore, for lubrication flows in narrow geometries with $\epsilon \ll1$, the Deborah number can be small but $Wi=O(1)$. In addition to the Deborah number $De=\lambda q/(2\ell h_{\ell})$ based on the exit height, we can introduce the Deborah number $De_{\it entry}=\lambda q/(2\ell h_{0})$ based on the entry height; the two Deborah numbers are related through $De_{\it entry}=De/H_0$. 

\subsection{Non-dimensional governing equations in Cartesian coordinates}

Substituting the non-dimensional variables (\ref{Scaling variables FCR})--(\ref{De_Wi FCR}) into the governing equations (\ref{Continuity+Momentum FCR})--(\ref{Conformation tensor FCR}) and considering the leading order in $\epsilon$, we obtain 
\begin{subequations}\label{Non-dim Gov.Eq FCR}
\begin{equation}\label{Non-dim continuity FCR}
\frac{\partial U_z}{\partial Z}+\frac{\partial U_y}{\partial Y}=0,
\end{equation}
\begin{equation}\label{Non-dim.z-momentum FCR}
\frac{\partial P}{\partial Z}=(1-\tilde\beta)\frac{\partial^2 U_z}{\partial Y^2}+\frac{\tilde\beta}{De}\left(\frac{\partial (\mathcal{F}(\tilde{\mathsfbi{A}})\tilde A_{zz})}{\partial Z}+ \frac{\partial (\mathcal{F}(\tilde{\mathsfbi{A}})\tilde A_{yz})}{\partial Y}\right),
\end{equation}
\begin{equation}\label{Non-dim y-momentum FCR}
\frac{\partial P}{\partial Y}=0,
\end{equation}
\begin{equation}\label{Non-dim Azz FCR}
U_z\frac{\partial \tilde A_{zz}}{\partial Z}+U_y\frac{\partial \tilde A_{zz}}{\partial Y}-2\frac{\partial U_z}{\partial Z}\tilde A_{zz}-2\frac{\partial U_z}{\partial Y}\tilde A_{yz}=-\frac{\mathcal{F}(\tilde{\mathsfbi{A}})}{De}\tilde{A}_{zz},
\end{equation}
\begin{equation}\label{Non-dim Ayz FCR} 
U_z\frac{\partial \tilde A_{yz}}{\partial Z}+U_y\frac{\partial \tilde A_{yz}}{\partial Y}-\frac{\partial U_y}{\partial Z}\tilde A_{zz}-\frac{\partial U_z}{\partial Y}\tilde A_{yy}=-\frac{\mathcal{F}(\tilde{\mathsfbi{A}})}{De}\tilde{A}_{yz},
\end{equation}
\begin{equation}\label{Non-dim Ayy FCR}
U_z\frac{\partial \tilde A_{yy}}{\partial Z}+U_y\frac{\partial \tilde A_{yy}}{\partial Y}-2\frac{\partial U_y}{\partial Z}\tilde A_{yz}-2\frac{\partial U_y}{\partial Y}\tilde A_{yy}=-\frac{\mathcal{F}(\tilde{\mathsfbi{A}})}{De}(\tilde{A}_{yy}-1),
\end{equation}
\end{subequations}
where
\begin{equation}\label{Non-dim Warner spring FCR}
\mathcal{F}(\tilde{\mathsfbi{A}})=\frac{1}{1-\dfrac{1}{\epsilon^2 L^2}(\tilde{A}_{zz}+\epsilon^2 \tilde{A}_{yy})}\approx\frac{1}{1-\tilde{A}_{zz}/(\epsilon^2 L^2)}+O(\epsilon^2).
\end{equation}
From the $y$-momentum equation, (\ref{Non-dim Gov.Eq FCR}$c$), it follows that $P=P(Z)+O(\epsilon^2)$, i.e. the pressure is constant across a cross-section but varies along the $z$-direction.  
Under the non-dimensionalization (\ref{Scaling A_zz, A_yz, A_yy}), the right-hand side of (\ref{Non-dim Gov.Eq FCR}$d$) becomes $-(\mathcal{F}(\tilde{\mathsfbi{A}})/De)(\tilde{A}_{zz}-\epsilon^2)$. Thus, at the leading order in $\epsilon$, we have $-(\mathcal{F}(\tilde{\mathsfbi{A}})/De)\tilde{A}_{zz}$.

For lubrication flows through the slowly varying geometries that we consider, (\ref{Non-dim Warner spring FCR}) clearly indicates that the finite extensibility is governed by the dimensionless parameter $\epsilon^2 L^2$ rather than $L^2$~\citep{ahmed2023modeling,housiadas2023_2D}. Although we consider $\epsilon \ll 1$, since the realistic values of $L^2$ are typically large~(see, e.g.,~\cite{remmelgas1999computational,rothstein1999extensional}),  we may have $\epsilon^2 L^2=O(1)$.

The corresponding boundary conditions on the velocity are
\refstepcounter{equation}
$$
 U_{z}(H(Z),Z)=0, \quad U_{y}(H(Z),Z)=0,  \quad \frac{\partial U_{z}}{\partial Y}(0,Z)=0,  \quad \int^{H(Z)}_0 U_{z}(Y,Z)\mathrm{d}Y=1. \eqno{(\theequation{a\mbox{$-$}d})} \label{ND BC FCR Velocity}
$$
These boundary conditions represent, respectively, the no-slip and
no-penetration conditions along the channel walls, the symmetry boundary condition at the centerline, and the integral mass conservation along the channel. 
In addition, we assume a fully developed unidirectional flow of a FENE-CR fluid in the straight entry channel, given by the Poiseuille velocity profile, and the corresponding conformation tensor (see the derivation in Appendix~\ref{App: straight channel})
\begin{subequations}
\begin{equation}
\tilde A_{zz}=L^2\epsilon^2+L^3\epsilon^3\frac{L\epsilon-\sqrt{L^2\epsilon^2+72De^2 Y^2/H^6_0}}{36 De^2 Y^2/H^6_0},
\end{equation}
\begin{equation}
\tilde{A}_{yz}=L\epsilon\frac{L\epsilon-\sqrt{L^2\epsilon^2+72De^2 Y^2/H^6_0}}{12De Y/H^3_0} \quad \text{and} \quad  \tilde{A}_{yy}=1.
\end{equation}
\label{ND BC FCR A}\end{subequations}

\subsection{Non-dimensional pressure drop across the contraction}\label{dP section CartC}

The integral mass conservation along the channel (\ref{ND BC FCR Velocity}$d$) sets the local value of the pressure gradient and allows one to calculate the pressure drop without solving for the velocity field. Integrating by parts the integral constraint
(\ref{ND BC FCR Velocity}$d$) and using (\ref{ND BC FCR Velocity}$a$) and (\ref{ND BC FCR Velocity}$c$), we obtain~\citep{BoykoHinchStone2023,HinchBoykoStone2023}
\begin{equation}
1=\int^{H(Z)}_0 U_{z}\mathrm{d}Y=-\int_{0}^{H(Z)}Y\frac{\partial U_z}{\partial Y}\mathrm{d}Y=-\frac{1}{2}\int_{0}^{H(Z)}(H(Z)^2-Y^{2})\frac{\partial^{2}U_z}{\partial Y^{2}}\mathrm{d}Y.\label{Integration by parts, repeatedly}
\end{equation}
Substituting the expression for $\partial^{2}U_z/\partial Y^{2}$ from the momentum equation (\ref{Non-dim Gov.Eq FCR}$b$) into (\ref{Integration by parts, repeatedly}) and rearranging provides an expression for the pressure gradient,
\begin{equation}
\frac{\mathrm{d}P}{\mathrm{d}Z}=-\frac{3(1-\tilde{\beta})}{H(Z)^{3}}+\frac{3 \tilde{\beta}}{2De H(Z)^{3}}\int_{0}^{H(Z)}(H(Z)^2-Y^{2})\left[\frac{\partial (\mathcal{F}(\tilde{\mathsfbi{A}})\tilde A_{zz})}{\partial Z}+ \frac{\partial (\mathcal{F}(\tilde{\mathsfbi{A}})\tilde A_{yz})}{\partial Y}\right]\mathrm{d}Y.\label{dP/dZ expl}
\end{equation}
Next, integrating (\ref{dP/dZ expl}) with respect to $Z$ from 0
to 1 yields the pressure drop $\Delta  P=P(0)-P(1)$ across the non-uniform region
\begin{eqnarray}
&&\mbox{\hspace{-7mm}} \Delta P=3(1-\tilde{\beta}) \int_{0}^{1} \frac{\mathrm{d} Z}{H(Z)^{3}} \nonumber\\
&&\mbox{\hspace{-7mm}}-\frac{3 \tilde{\beta}}{2De }\int_{0}^{1}\frac{1}{ H(Z)^{3}}\int_{0}^{H(Z)}(H(Z)^2-Y^{2})\left[\frac{\partial (\mathcal{F}(\tilde{\mathsfbi{A}})\tilde A_{zz})}{\partial Z}+ \frac{\partial (\mathcal{F}(\tilde{\mathsfbi{A}})\tilde A_{yz})}{\partial Y}\right]\mathrm{d}Y \mathrm{d} Z.
\label{dP 1 general}
\end{eqnarray}
Finally, using integration by parts, (\ref{dP 1 general}) can be expressed as
\begin{eqnarray}
&&\mbox{\hspace{-2mm}} \Delta P= (1-\tilde{\beta})\Delta\hat{P}+\frac{\tilde{\beta}}{De }\int_{0}^{H(0)}[\mathcal{F}(\tilde{\mathsfbi{A}})\tilde A_{zz}\hat{U}_{z}]_{Z=0}\mathrm{d}Y-\frac{\tilde{\beta}}{De}\int_{0}^{H(1)}[\mathcal{F}(\tilde{\mathsfbi{A}})\tilde A_{zz}\hat{U}_{z}]_{Z=1}\mathrm{d}Y \nonumber\\
&&+\frac{\tilde{\beta}}{De }\int_{0}^{1}\int_{0}^{H(Z)}\mathcal{F}(\tilde{\mathsfbi{A}})\tilde A_{zz}\frac{\partial\hat{U}_{z}}{\partial Z}\mathrm{d}Y\mathrm{d}Z+\frac{\tilde{\beta}}{De }\int_{0}^{1}\int_{0}^{H(Z)}\mathcal{F}(\tilde{\mathsfbi{A}})\tilde A_{yz}\frac{\partial\hat{U}_{z}}{\partial Y}\mathrm{d}Y\mathrm{d}Z.
\label{dP non-uniform by parts}
\end{eqnarray}
Here, the function $\mathcal{F}(\tilde{\mathsfbi{A}})$ is given in (\ref{Non-dim Warner spring FCR}), and $\Delta\hat{P}$ and $\hat{U}_{z}$ are the corresponding pressure drop and axial velocity of a Newtonian fluid given by~\citep{boyko2022pressure}
\refstepcounter{equation}
$$
\Delta \hat{P}=3\int_{0}^{1}\frac{\mathrm{d}Z}{H(Z)^{3}}, \qquad \hat{U}_{z}=\frac{3}{2}\frac{H(Z)^2-Y^2}{H(Z)^3}. \eqno{(\theequation{{ a,b}})}\label{Newtonian dP Uz}
$$
Equation (\ref{dP non-uniform by parts}) represents
the expression for the non-dimensional pressure drop previously obtained from an application of the reciprocal theorem in a slowly varying channel~\citep{boyko2021RT,boyko2022pressure}. The first term on the right-hand side of (\ref{dP non-uniform by parts}) represents the contribution of the Newtonian solvent to the pressure drop. The second and third terms represent the contribution of the elastic normal stresses at the inlet and outlet of the non-uniform channel. Finally, the fourth and fifth terms represent the contribution of the elastic normal stresses and elastic shear stresses within the non-uniform channel.

\section{Low-Deborah-number lubrication analysis}\label{Low-Deborah-number lubrication analysis}

In this section, we employ the low-Deborah-number lubrication analysis to derive asymptotic expressions for the velocity, conformation
tensor, and pressure drop of a weakly viscoelastic FENE-CR fluid up to $O(De^4)$.
To this end, we expand the velocity, pressure drop, and conformation tensor components into perturbation series in the Deborah number $De\ll1$,
\begin{equation}\label{perturbation_expansion FCR}
    \begin{pmatrix}
 U_z\\
 U_y\\
 P\\
 \tilde A_{zz}\\
 \tilde A_{yy}\\
 \tilde A_{yz}
 \end{pmatrix} = \begin{pmatrix}
 U_{z,0}\\
 U_{y,0}\\
 P_0\\
 \tilde A_{zz,0}\\
\tilde A_{yy,0}\\
 \tilde A_{yz,0}
 \end{pmatrix}+De \begin{pmatrix}
 U_{z,1}\\
 U_{y,1}\\
 P_1\\
 \tilde A_{zz,1}\\
 \tilde A_{yy,1}\\
 \tilde A_{yz,1}
 \end{pmatrix}+De^2\begin{pmatrix}
 U_{z,2}\\
 U_{y,2}\\
 P_2\\
 \tilde A_{zz,2}\\
 \tilde A_{yy,2}\\
 \tilde A_{yz,2}
 \end{pmatrix}+... \:.
\end{equation}
As noted by~\citet{boyko2022pressure}, in the weakly viscoelastic and lubrication limits, $De\ll1$ and $\epsilon \ll1$, it is sufficient to apply the boundary conditions on the velocity (\ref{ND BC FCR Velocity}) to find the flow field, conformation tensor components, and pressure drop at each order in $De$. 
Indeed, the iterative structure of the solution eliminates the need to use the boundary condition (\ref{ND BC FCR A}) on the conformation tensor~\citep{black1976converging,boyko2022pressure,housiadas2023_2D}. 
For example, considering the leading and first order in $De$, we find
\begin{equation} 
\tilde A_{zz,0}=0,\qquad \tilde A_{yz,0}=0,\qquad \tilde A_{yy,0}=1, \label{Leading-order Azz Ayz Ayy FCR}
\end{equation} 
\begin{equation} 
\tilde A_{zz,1}=0,\qquad \tilde A_{yz,1}=\frac{\partial U_{z,0}}{\partial Y}, \qquad \tilde A_{yy,1}=2\frac{\partial U_{y,0}}{\partial Y}. \label{First-order Azz Ayz Ayy FCR} 
\end{equation} 

In Appendix~\ref{Details of low De}, we provide a detailed derivation of the expressions for the pressure drop of the FENE-CR fluid in the low-$De$ limit up to $O(De^4)$. We obtain that the expressions for the pressure drop at the leading, first, and second order in $De$ are the same for the FENE-CR and Oldroyd-B fluids, and are given by 
\refstepcounter{equation}
$$
 \Delta P_{0}=3\int_{0}^{1}\frac{\mathrm{d}Z}{H(Z)^{3}}, \qquad \Delta P_1 = \frac{9}{2}\tilde \beta \left(\frac{1}{H(0)^4}-\frac{1}{H(1)^4} \right),\eqno{(\theequation{a,b})}\label{Leading, first, second order sol FCR}
$$
$$
  \Delta P_2=\frac{324}{35}\tilde\beta \int^1_0\left(\frac{14 H'(Z)^2}{H(Z)^7} - \frac{3 H''(Z)}{H(Z)^6} \right) \mathrm{d}Z. \eqno{(\theequation{c})}
$$Interestingly, unlike the FENE-CR fluid, the pressure drop of the FENE-P fluid is different from the Oldroyd-B case at $O(De^2)$ and depends on finite extensibility through $L^2\epsilon^2$, as recently shown by~\citet{housiadas2023_2D}.

At the third order in $De$, the pressure drop of the FENE-CR fluid is different from the Oldroyd-B fluid due to the finite extensibility and is given as
\begin{eqnarray}
 \Delta P_{3}&=&- \frac{2673\tilde{\beta}}{70L^2\epsilon^2}\left (\frac{1}{H(0)^{8}}-\frac{1}{H(1)^{8}}\right )  \nonumber\\
&+&\frac{648\tilde{\beta}(9-\tilde{\beta})}{35}\left(\frac{H'(0)^{2}}{H(0)^{8}}-\frac{H'(1)^{2}}{H(1)^{8}}\right)-\frac{216\tilde{\beta}(8-\tilde{\beta})}{35}\left (\frac{H''(0)}{H(0)^{7}}-\frac{H''(1)}{H(1)^{7}}\right ).
\label{pressure drop sol third order FCR}
\end{eqnarray}
From (\ref{pressure drop sol third order FCR}), it follows that $\Delta P_{3}$ may increase, decrease, or not change the total pressure drop of the FENE-CR fluid, depending on the geometry. 
For a contraction ($H(0)>H(1)$), the first term, which depends on finite extensibility through $L^2\epsilon^2$ and distinguishes the FENE-CR fluid from the Oldroyd-B fluid, leads to an increase in the pressure drop. However, for an expansion ($H(0)<H(1)$), the first term leads to a decrease in the pressure drop, and for a constriction ($H(1)=H(0)$) it does not contribute to the pressure drop. We also note that our expression for the pressure drop $\Delta P_3$ of the FENE-CR fluid is similar to the expression for the pressure drop of the FENE-P fluid at $O(De^3)$, albeit a different number in the coefficient of the first term in (\ref{pressure drop sol third order FCR})~\citep{housiadas2023_2D}.

Finally, at the fourth order in $De$, the resulting expression for the pressure drop of the FENE-CR fluid is
\begin{eqnarray}\label{pressure drop sol fourth order FCR}
     \Delta P_4&=&\frac{3888\tilde\beta (8 \tilde\beta+25)}{175L^2 \epsilon ^2}\left[\frac{H'(1)}{H(1)^{10}}-\frac{H'(0)}{H(0)^{10}} \right]+\frac{648 \tilde\beta}{175L^2 \epsilon ^2}\int_0^1\left[a_{1}\frac{H''}{H^{10}} +a_{2}\frac{H'^2}{H^{11}} \right]\mathrm{d}Z \nonumber \\
     &&+\int_0^1 \left[a_3 \frac{H''^2}{H^9} +a_4\frac{H''' H'}{H^9}+a_5\frac{H^3 H''''}{H^{11}}+ a_6\frac{H'^4}{H^{11}} + a_7  \frac{H'^2 H''}{H^{10}}  \right]\mathrm{d}Z\nonumber \\
     &&+a_8\left[\frac{H'''(0)}{H(0)^8}-\frac{H'''(1)}{H(1)^8}   \right]+a_9\left[\frac{H'(1) H''(1)}{H(1)^9}-\frac{H'(0) H''(0)}{H(0)^9}\right]\nonumber \\
     &&+a_{10}\left[\frac{H'(0)^3}{H(0)^{10}}-\frac{H'(1)^3} {H(1)^{10}}\right] ,
 \end{eqnarray}
where the coefficients $a_1, . . ., a_{10}$ are summarized in table~\ref{T1}.

\begin{table}
  \begin{center}
\def~{\hphantom{0}}
   \begin{tabular}{cccc}
Coefficient    &   Expression & Coefficient   &  Expression\\
\hline
\noalign{\smallskip}

$a_1$ &   $\displaystyle{55-8 \tilde\beta}$ &   $a_2$ & $\displaystyle{20(4\tilde\beta-15)}$ \\ [1.5ex]

$a_3$ &   $\displaystyle{\frac{3240 \tilde\beta }{13475 } [\tilde\beta (41-70 \tilde\beta)+910]}$ &   $a_4$ & $\displaystyle{\frac{4536 \tilde\beta }{13475 }[\tilde\beta (119-82 \tilde\beta)+750]}$ \\ [1.5ex]
 
$a_5$ &     $\displaystyle{\frac{1296 \tilde\beta }{13475} [2 \tilde\beta (7 \tilde\beta-5)-175]}$ & $a_6$ &  $\displaystyle{\frac{9072 \tilde\beta }{13475} [11 \tilde\beta (83-40 \tilde\beta)+2400]}$   \\[1.5ex]

$a_7$  &    $\displaystyle{\frac{1944 \tilde\beta }{13475}[\tilde\beta (1666 \tilde\beta-2789)-12950]}$  & $a_8$ & $\displaystyle{\frac{5184 \tilde\beta}{13475} [3 \tilde\beta (7 \tilde\beta-24)+175]}$ \\[1.5ex]

$a_9$ & $\displaystyle{\frac{2592 \tilde\beta }{13475}[3 \tilde\beta (175 \tilde\beta-618)+4550]}$ & $a_{10}$ & $\displaystyle{\frac{3888 \tilde\beta }{1925}[11 \tilde\beta (8 \tilde\beta-29)+800]}$   \\[1.5ex]
\end{tabular}
  \caption{Coefficients appearing in the expression (\ref{pressure drop sol fourth order FCR}) for the fourth-order pressure drop $\Delta P_4$ of the FENE-CR fluid in a planar contracting channel.}
  \label{T1}
  \end{center}
\end{table}

The first two terms on the right-hand side of (\ref{pressure drop sol fourth order FCR}) depend on $L^2\epsilon^2$, and thus clearly distinguish the analytical prediction for $\Delta P_4$ of the FENE-CR fluid from the Oldroyd-B fluid. For the Oldroyd-B fluid, our analytical result for $\Delta P_4$ fully agrees with the solution of~\citet{housiadas2023_2D} when accounting for the differences in the non-dimensionalization. However, as expected based on the previous orders, our expression for the fourth-order pressure drop of the FENE-CR fluid differs from the expression for the FENE-P fluid given in~\cite{housiadas2023_2D}. Specifically, the first two terms in (\ref{pressure drop sol fourth order FCR}) that include $L^2\epsilon^2$ appear in the fourth-order expressions for both FENE-CR and FENE-P fluids with different coefficients. Furthermore, the expression for the FENE-P fluid has an additional term of the form of $\int^1_0 H(Z)^{-11} \mathrm{d}Z$ that depends on $1/L^4\epsilon^4$. 

For a given flow rate, we have determined the dimensionless pressure
drop $\Delta P=\Delta p/(\mu_{0}q\ell/2h_{\ell}^{3})$ of a FENE-CR fluid as a function
of the shape function $H(Z)$, the viscosity ratio $\tilde{\beta}$, the parameter $L^2\epsilon^2$,  and the
Deborah number $De$ up to $O(De^4)$,
\begin{equation}
\Delta P=\Delta P_{0}+De\Delta P_{1}+De^{2}\Delta P_{2}+De^{3}\Delta P_{3}+De^{4}\Delta P_{4}+O(\epsilon^{2},De^{5}),\label{Pressure drop RT 2D}
\end{equation}
where the expressions for $\Delta P_{0}$, $\Delta P_{1}$, $\Delta P_{2}$, $\Delta P_{3}$, and $\Delta P_{4}$
are given in (\ref{Leading, first, second order sol FCR}$a$), (\ref{Leading, first, second order sol FCR}$b$), (\ref{Leading, first, second order sol FCR}$c$),
(\ref{pressure drop sol third order FCR}), and (\ref{pressure drop sol fourth order FCR}), respectively.
Physically, the non-dimensional quantity $\Delta P=\Delta p/(\mu_{0}q\ell/2h_{\ell}^{3})$ represents
the dimensionless flow resistance ($\Delta p/q$) for a given geometry.

Having the low-$De$ asymptotic expressions for $\Delta P_{0}$, $\Delta P_{1}$, $\Delta P_{2}$, $\Delta P_{3}$, and $\Delta P_{4}$, we can improve the convergence of the asymptotic series (\ref{Pressure drop RT 2D}) by using the diagonal Pad\'{e} [2/2] approximation~\citep{hinchperturbation,housiadas2017improved,housiadas2023_2D}, 
\begin{equation}
\Delta P_{ \it Pade}=\Delta P_{0}+De\frac{De(\Delta P_{2})^{3}+\Delta P_{1}\Delta P_{2}(\Delta P_{2}-2De\Delta P_{3})+\Delta P_{1}^{2}(De\Delta P_{4}-\Delta P_{3})}{(\Delta P_{2})^{2}+De^{2}(\Delta P_{3})^{2}+\Delta P_{1}(De\Delta P_{4}-\Delta P_{3})-De\Delta P_{2}(De\Delta P_{4}+\Delta P_{3})}.\label{Pade approximation dP}
\end{equation} 
It should be noted that~\citet{housiadas2023_2D} extended the low-Deborah-number lubrication analysis to much higher asymptotic orders and provided analytical expressions for the pressure drop of the Oldroyd-B and FENE-P fluids up to $O(De^8)$. Nevertheless, as shown for the Oldroyd-B fluid, the low-$De$ perturbation solutions obtained from the Pad\'{e} approximations remain indistinguishable when adding more terms in the asymptotic series beyond $O(De^4)$. 

\section{Low-\texorpdfstring{$\tilde{\beta}$}{} lubrication analysis}\label{Low-beta analysis-NU region}

In the previous section, we have derived analytical expressions for the non-dimensional pressure drop of a FENE-CR fluid in a non-uniform channel of arbitrary shape $H(Z)$ in the low-Deborah-number limit, $De\ll1$.  However, as pointed out by~\citet{BoykoHinchStone2023} and~\citet{HinchBoykoStone2023}, the low-Deborah-number asymptotic analysis cannot accurately predict the pressure drop at high $De$ numbers where there are significant elastic stresses. 

In this section, we employ orthogonal curvilinear coordinates and consider the ultra-dilute limit, $\tilde{\beta}=\mu_{p}/\mu_{0}\ll1$~\citep{remmelgas1999computational,moore2012weak,li2019orientation,mokhtari2022birefringent,BoykoHinchStone2023,HinchBoykoStone2023}, which allows us to analyze the pressure drop and conformation tensor at high Deborah numbers.

\subsection{Orthogonal curvilinear coordinates for a slowly varying geometry}\label{SecOCC}

For our low-$\tilde{\beta}$ lubrication analysis, we first transform the geometry of the contraction
from the Cartesian coordinates ($Z,Y$) to orthogonal curvilinear coordinates ($\xi,\eta$) with the mapping~\citep{BoykoHinchStone2023,HinchBoykoStone2023}
\begin{equation}
\xi=Z-\frac{1}{2}\epsilon^{2}\frac{H'(Z)}{H(Z)}(H(Z)^{2}-Y^{2})+O(\epsilon^{4}), \qquad\eta=\frac{Y}{H(Z)},\label{Direct transform paper}
\end{equation}
and use $\boldsymbol{u}=u\boldsymbol{e}_{\xi}+\varv\boldsymbol{e}_{\eta}$ and $\mathsfbi{A}=A_{11}\boldsymbol{e}_{\xi}\boldsymbol{e}_{\xi}+A_{12}(\boldsymbol{e}_{\xi}\boldsymbol{e}_{\eta}+\boldsymbol{e}_{\eta}\boldsymbol{e}_{\xi})+A_{22}\boldsymbol{e}_{\eta}\boldsymbol{e}_{\eta}$  to denote the components of velocity and conformation tensor in curvilinear coordinates ($\xi,\eta$).

The corresponding components of the non-dimensional velocity field and conformation tensor in different coordinates are related through
\begin{subequations}
\begin{equation}
U_{z}=U-\epsilon^{2}\eta H'(\xi) V, \qquad  U_{y}=\eta H'(\xi) U+V,\label{Relation between velocity ND}
\end{equation}
\begin{equation}
\tilde{A}_{zz}=\tilde{A}_{11}+O(\epsilon^{2}),\label{Relation between A ND 1}
\end{equation}
   \begin{equation}
\tilde{A}_{zy}=\tilde{A}_{12}+\eta H'(\xi)\tilde{A}_{11}+O(\epsilon^{2}),\label{Relation between A ND 2}
\end{equation}
\begin{equation}
\tilde{A}_{yy}=\tilde{A}_{22}+2\eta H'(\xi)\tilde{A}_{12}+\eta^2 (H'(\xi))^2\tilde{A}_{11}+O(\epsilon^{2}).\label{Relation between A ND 3}
\end{equation}\label{Relation between A ND paper}\end{subequations}
Note that, since there is only a $O(\epsilon^{2})$ difference between the $\xi$- and $z$-directions, for convenience, we prefer to use $Z$ rather than $\xi$ in curvilinear coordinates~\citep{BoykoHinchStone2023}.

\subsection{Non-dimensional governing equations in orthogonal curvilinear coordinates}

Using the mapping (\ref{Direct transform paper}), the governing equations (\ref{Non-dim Gov.Eq FCR})--(\ref{Non-dim Warner spring FCR}) and the corresponding boundary conditions (\ref{ND BC FCR Velocity})--(\ref{ND BC FCR A})  in curvilinear coordinates~\citep{BoykoHinchStone2023,HinchBoykoStone2023} take the form 
\begin{subequations}
\begin{equation}
\frac{\partial(HU)}{\partial Z}+\frac{\partial V}{\partial\eta}=0,\label{Continuity ND CC}
\end{equation}
\begin{equation}
\frac{\mathrm{d}P}{\mathrm{d}Z}=(1-\tilde{\beta})\frac{1}{H^{2}}\frac{\partial^{2}U}{\partial\eta^{2}}+\frac{\tilde{\beta}}{De}\left(\frac{1}{H}\frac{\partial(H\mathcal{F}(\tilde{\mathsfbi{A}})\tilde{A}_{11})}{\partial Z}+\frac{1}{H}\frac{\partial(\mathcal{F}(\tilde{\mathsfbi{A}})\tilde{A}_{12})}{\partial\eta}\right),\label{Momentum z CC}
\end{equation}
\begin{equation}
U\frac{\partial\tilde{A}_{11}}{\partial Z}+\frac{V}{H}\frac{\partial\tilde{A}_{11}}{\partial\eta}-2\frac{\partial U}{\partial Z}\tilde{A}_{11}-\frac{2}{H}\frac{\partial U}{\partial\eta}\tilde{A}_{12}=-\frac{\mathcal{F}(\tilde{\mathsfbi{A}})}{De}\tilde{A}_{11},\label{A_11 ND CC}
\end{equation}
\begin{equation}
U\frac{\partial\tilde{A}_{12}}{\partial Z}+\frac{V}{H}\frac{\partial\tilde{A}_{12}}{\partial\eta}-H\frac{\partial}{\partial Z}\left(\frac{V}{H}\right)\tilde{A}_{11}-\frac{1}{H}\frac{\partial U}{\partial\eta}\tilde{A}_{22}=-\frac{\mathcal{F}(\tilde{\mathsfbi{A}})}{De}\tilde{A}_{12},\label{A_12 ND CC}
\end{equation}
\begin{equation}
U\frac{\partial\tilde{A}_{22}}{\partial Z}+\frac{V}{H}\frac{\partial\tilde{A}_{22}}{\partial\eta}-2H\frac{\partial}{\partial Z}\left(\frac{V}{H}\right)\tilde{A}_{12}+2\frac{\partial U}{\partial Z}\tilde{A}_{22}=-\frac{\mathcal{F}(\tilde{\mathsfbi{A}})}{De}(\tilde{A}_{22}-1),\label{A_22 ND CC}
\end{equation}\label{ND CC}\end{subequations}
where
\begin{equation}
\mathcal{F}(\tilde{\mathsfbi{A}})=\frac{1}{1-\dfrac{1}{\epsilon^2 L^2}( \tilde{A}_{11}+\epsilon^2 \tilde{A}_{22})}\approx\frac{1}{1-\tilde{A}_{11}/(\epsilon^2 L^2)}, \label{Non-dim Warner spring curve}
\end{equation}
subject to the boundary conditions
\refstepcounter{equation}
$$
 U(Z,1)=0,\quad V(Z,1)=0,\quad \frac{\partial U} {\partial\eta}(Z,0)=0,\quad H(Z)\int_{0}^{1}U(Z,\eta)\mathrm{d}\eta=1, \eqno{(\theequation{a\mbox{$-$}d})}\label{ND BC CC}
$$
and
\begin{subequations}
\begin{equation}
\tilde A_{11}(0,\eta)=L^2\epsilon^2+L^3\epsilon^3\frac{L\epsilon-\sqrt{L^2\epsilon^2+72De^2 \eta^2/H^4_0}}{36 De^2 \eta^2/H^4_0},\label{A_11 BC entrance}
\end{equation}
\begin{equation}
\tilde{A}_{12}(0,\eta)=L\epsilon\frac{L\epsilon-\sqrt{L^2\epsilon^2+72De^2 \eta^2/H^4_0}}{12De \eta/H^2_0} \quad \text{and} \quad \tilde{A}_{22}(0,\eta)=1. \label{A_12 BC entrance}
\end{equation}
\label{ND BC CC A}\end{subequations}
Following similar steps as in $\mathsection$~\ref{dP section CartC} and using the integral constraint (\ref{ND BC CC}$d$), the non-dimensional pressure drop can be expressed in curvilinear coordinates as
\begin{eqnarray}
\mbox{\hspace{-4mm}}&&\Delta P= \underset{\mathrm{Solvent\,stress\,contribution}}{\underbrace{3(1-\tilde{\beta})\int_{0}^{1}\frac{\mathrm{d}Z}{H(Z)^{3}}}}
+\underset{\mathrm{Elastic\,shear\,stress\,contribution}}{\underbrace{-\frac{3\tilde{\beta}}{De}\int_{0}^{1}\left[\frac{1}{H(Z)}\int_{0}^{1}\eta\mathcal{F}(\tilde{\mathsfbi{A}})\tilde{A}_{12}\mathrm{d}\eta\right]\mathrm{d}Z}}
 \nonumber\\
&&\mbox{\hspace{-10mm}}+\underset{\mathrm{Elastic\,normal\,stress\,contribution}}{\underbrace{\frac{3\tilde{\beta}}{2De}\left(\int_{0}^{1}(1-\eta^{2})\left[\mathcal{F}(\tilde{\mathsfbi{A}})\tilde{A}_{11}\right]^{0}_{1}\mathrm{d}\eta-\int_{0}^{1}\left[\frac{H'(Z)}{H(Z)}\left(\int_{0}^{1}(1-\eta^{2})\mathcal{F}(\tilde{\mathsfbi{A}})\tilde{A}_{11}\mathrm{d}\eta\right)\right]\mathrm{d}Z\right)}},
\label{dP non-uniform by parts FENE-CR CC}
\end{eqnarray}
where $[\mathcal{F}(\tilde{\mathsfbi{A}})\tilde{A}_{11}]^{0}_{1}=\mathcal{F}(\tilde{\mathsfbi{A}})\tilde{A}_{11}|_{Z=0}-\mathcal{F}(\tilde{\mathsfbi{A}})\tilde{A}_{11}|_{Z=1}$.

Equation (\ref{dP non-uniform by parts FENE-CR CC}) represents the pressure drop in curvilinear coordinates and is an analog of (\ref{dP non-uniform by parts}), written in Cartesian coordinates. The first term on the right-hand side of
(\ref{dP non-uniform by parts FENE-CR CC}) represents the viscous contribution of the Newtonian solvent to the pressure drop. The second term represents the contribution of the elastic shear stresses and the last term represents the contribution of
the elastic normal stresses to the pressure drop.

\subsection{Velocity, conformation, and pressure drop in the ultra-dilute limit }


Next, we consider the ultra-dilute limit, $\tilde{\beta} \ll 1$,  representing a one-way coupling between the velocity and pressure
fields and the conformation tensor.
At the leading order in $\tilde{\beta}$, the velocity field of the FENE-CR fluid is parabolic, similar to Newtonian and Oldroyd-B fluids, and is given as
\refstepcounter{equation}
$$
U=\frac{3}{2}\frac{1}{H(Z)}(1-\eta^{2})\qquad\hbox{and}\qquad V\equiv0. \eqno{(\theequation{a,b})}\label{U0 and V0 2D curve}
$$
We note that in orthogonal curvilinear coordinates, the velocity in the $\eta$-direction is identically zero at $O(\tilde{\beta}^{0})$, in contrast to the Cartesian coordinates where $U_{y}=(3/2)H'(Z)Y(H(Z)^{2}-Y^{2})/H(Z)^{4}$. As pointed out by \citet{BoykoHinchStone2023}, the latter fact significantly simplifies the theoretical analysis. 

Substituting (\ref{U0 and V0 2D curve}) into (\ref{A_11 ND CC})$-$(\ref{A_22 ND CC}) and using (\ref{Non-dim Warner spring curve}), we obtain the simplified equations for the conformation tensor components of the FENE-CR fluid at leading order in $\tilde{\beta}$,
\begin{subequations}
\begin{equation}
U\frac{\partial\tilde{A}_{22}}{\partial Z}+2\frac{\partial U}{\partial Z}\tilde{A}_{22}=-\frac{1}{De}\frac{1}{1-\tilde{A}_{11}/(\epsilon^2 L^2)}(\tilde{A}_{22}-1),\label{A ND CC LOa} 
\end{equation}
\begin{equation}
U\frac{\partial\tilde{A}_{12}}{\partial Z}-\frac{1}{H}\frac{\partial U}{\partial\eta}\tilde{A}_{22}=-\frac{1}{De}\frac{1}{1-\tilde{A}_{11}/(\epsilon^2 L^2)}\tilde{A}_{12},\label{A ND CC LOb}
\end{equation}
\begin{equation}
U\frac{\partial\tilde{A}_{11}}{\partial Z}-2\frac{\partial U}{\partial Z}\tilde{A}_{11}-\frac{2}{H}\frac{\partial U}{\partial\eta}\tilde{A}_{12}=-\frac{1}{De}\frac{1}{1-\tilde{A}_{11}/(\epsilon^2 L^2)}\tilde{A}_{11},\label{A ND CC LOc}
\end{equation}\label{A ND CC LO}\end{subequations}
where $U$ is given in (\ref{U0 and V0 2D curve}$a$).

Equations (\ref{A ND CC LO}) represent a set of coupled first-order semi-linear partial
differential equations that should
be solved at once to obtain $\tilde{A}_{22}$, $\tilde{A}_{12}$, and $\tilde{A}_{11}$ for the FENE-CR fluid.
When $L^2\epsilon^2 \to \infty$, corresponding to the Oldroyd-B fluid, (\ref{A ND CC LO}) reduces to a set of one-way coupled equations, allowing us to derive semi-analytical expressions for the conformation tensor for arbitrary values of the Deborah number in the ultra-dilute limit~\citep{BoykoHinchStone2023}.
Furthermore,~\citet{BoykoHinchStone2023} and~\citet{HinchBoykoStone2023} provided analytical expressions for the conformation tensor and the pressure drop of the Oldroyd-B fluid in the high-Deborah-number limit. In particular, the pressure drop of the Oldroyd-B fluid across the non-uniform channel in the high-$De$ limit is
\begin{equation}
\Delta P=\underset{\mathrm{Solvent\,stress}}{\underbrace{3(1-\tilde{\beta})\int_{0}^{1}\frac{\mathrm{d}Z}{H(Z)^{3}}}}+\underset{\mathrm{Elastic\,shear\,stress}}{\underbrace{3\tilde{\beta}H_0^{-2}\int_{0}^{1}\frac{\mathrm{d}Z}{H(Z)}}}+\underset{\mathrm{Elastic\,normal\,stress}}{\underbrace{\frac{9}{5}\tilde{\beta}De(H_0^{-4}-H_{0}^{-2})}}\;\mathrm{for}\; De\gg1.\label{Pressure drop OB high-De}
\end{equation}
The coupling between equations in (\ref{A ND CC LO}) greatly complicates the analytical progress, particularly in the high-$De$ asymptotic limit for the FENE-CR fluid.
Nevertheless, examining the expressions in (\ref{A ND CC LO}), we observe that for a given value of $\eta\in[0,1]$, (\ref{A ND CC LO}) represent a set of first-order ordinary differential equations for $\tilde{A}_{22}$, $\tilde{A}_{12}$, and $\tilde{A}_{11}$ of the FENE-CR fluid. Therefore, we solve numerically the coupled equations (\ref{A ND CC LO}) subject to the boundary conditions (\ref{ND BC CC A}) using MATLAB’s routine~\texttt{ode45} 
and obtain the distribution of $\tilde{A}_{22}$, $\tilde{A}_{12}$, and $\tilde{A}_{11}$ in a contraction for different values of $De$ and $H_0$ in the limit of $\tilde{\beta} \ll 1$. Typical values of the grid
size are $\Delta Z  = 10^{-3}$ and  $\Delta \eta = 0.005$. Once $\tilde{A}_{11}$ and $\tilde{A}_{12}$ are determined, we use MATLAB’s routine~\texttt{trapz}
to calculate the pressure drop (\ref{dP non-uniform by parts FENE-CR CC}) in a contraction.

\section{Results}

In this section, we present our theoretical results for the pressure drop and elastic stresses of the FENE-CR fluid as developed in the previous sections. We also validate the predictions of our theoretical model against the two-dimensional numerical simulations with the finite-volume software OpenFOAM. The details of the numerical procedure are provided in Appendix~\ref{Details of numerical simulations using OpenFOAM}. 
For comparison and validation, in addition to the FENE-CR fluid, we show the results for the Oldroyd-B fluid.

As an illustrative example, we consider a hyperbolic contracting channel of the form
\begin{equation}
    H(Z)=\frac{H_{0}}{(H_{0}-1)Z+1}, \label{channel profile 2D}
\end{equation}
where $H_{0}=h_0/h_\ell$ is the ratio of the heights at the inlet and outlet; for the contracting geometry we have $H_0>1$. The present study focuses on the contraction ratio $H_{0}=h_0/h_\ell=4$.

\subsection{Pressure drop at low Deborah numbers}

In this subsection, we elucidate the pressure drop behavior of the FENE-CR fluid at low Deborah numbers using our analytical predictions and OpenFOAM simulation results. In addition, we present the pressure drop of the Oldroyd-B and FENE-P fluids, thus highlighting how the finite extensibility (without the influence of shear thinning) incorporated by the FENE-CR model impacts pressure drop.

For the planar hyperbolic contracting channel (\ref{channel profile 2D}), using (\ref{Leading, first, second order sol FCR}$a$), (\ref{Leading, first, second order sol FCR}$b$), (\ref{Leading, first, second order sol FCR}$c$),
(\ref{pressure drop sol third order FCR}), and (\ref{pressure drop sol fourth order FCR}), we obtain analytical expressions for the pressure drop contributions of the FENE-CR fluid up to $O(De^4)$ 
\begin{subequations}\label{dP_FCR}
    \begin{equation}
    \Delta P_0=\frac{3}{4}(1+H^{-1}_{0}) (1+H^{-2}_{0}),\label{dP0_FCR}
        \end{equation}
        \begin{equation}
        \Delta P_1=-\frac{9}{2}\tilde\beta(1-H^{-4}_{0}),\label{dP1_FCR}
         \end{equation}
        \begin{equation}
        \Delta P_2=\frac{648}{35}\tilde\beta(1-H^{-1}_{0})^2(1+H^{-1}_{0})(1+H^{-2}_{0}),\label{dP2_FCR}
         \end{equation}
        \begin{equation}
        \Delta P_3=\frac{2673}{70L^2 \epsilon ^2}\tilde\beta(1-H^{-8}_{0})-\frac{216}{35}\tilde\beta(11-\tilde\beta)(1-H^{-1}_{0})^3(1+H^{-1}_{0})(1+H^{-2}_{0}),\label{dP3_FCR}
         \end{equation}
         \begin{eqnarray}
         \Delta P_4&=&-\frac{162}{35L^2 \epsilon ^2}\tilde\beta(32 \tilde\beta+139)(1-H^{-1}_{0}-H^{-8}_{0} +H^{-9}_{0}) \nonumber \\
        &&+\frac{324}{13475}\tilde\beta (840 \tilde\beta^2-3351 \tilde\beta+9800)(1-H^{-1}_{0})^4(1+H^{-1}_{0}(1+H^{-2}_{0}).\label{dP4_FCR}\end{eqnarray}\end{subequations}
Using (\ref{dP_FCR}) in conjunction with (\ref{Pade approximation dP}), we obtain the Pad\'{e} approximation for the pressure drop. Note that for $L^2\epsilon^2 \to \infty$, we recover the Oldroyd-B limit. In this case, the first terms in (\ref{dP3_FCR}) and (\ref{dP4_FCR}), which are dependent on $L^2\epsilon^2$, vanish.
\begin{figure}
 \centerline{\includegraphics[scale=1]{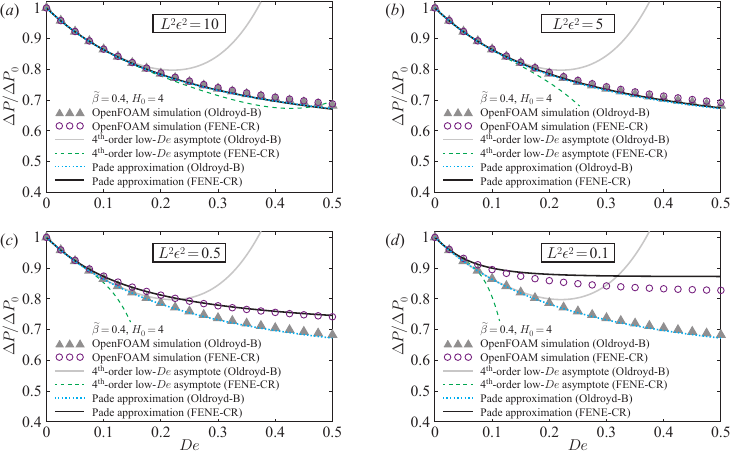}}
\caption{Non-dimensional pressure drop at low Deborah numbers for the Oldroyd-B and FENE-CR fluids in a contracting channel
described by~(\ref{channel profile 2D}). ($a$--$d$) Scaled pressure drop $\Delta P/\Delta P_0$ as a function of $De=\lambda q/(2\ell h_{\ell})$ for ($a$) $L^2\epsilon^2=10$, ($b$) $L^2\epsilon^2=5$, ($c$) $L^2\epsilon^2=0.5$, and ($d$) $L^2\epsilon^2=0.1$. Gray triangles and purple circles represent, respectively, the OpenFOAM simulation results for the Oldroyd-B and FENE-CR fluids. Gray solid and green dashed lines represent the fourth-order asymptotic solutions for the Oldroyd-B and FENE-CR fluids, given by (\ref{dP0_FCR})$-$(\ref{dP4_FCR}). Cyan dotted and solid black lines represent the Pad\'{e} approximation (\ref{Pade approximation dP}) applied to the fourth-order asymptotic solutions for the Oldroyd-B and FENE-CR fluids. All calculations were performed using $H_0=4$ and $\tilde\beta=0.4$.}\label{Fig:2}
\end{figure}

We present in figure~\ref{Fig:2} the scaled pressure drop $\Delta P/\Delta P_0$ as a function of $De=\lambda q/(2\ell h_{\ell})$ for the Oldroyd-B and FENE-CR fluids in a contracting channel for different values of $L^2\epsilon^2$. Gray triangles and purple circles represent the OpenFOAM simulation results for the Oldroyd-B and FENE-CR fluids obtained from calculating the pressure drop along the centreline. Gray solid and green dashed lines represent the fourth-order asymptotic solutions for the Oldroyd-B and FENE-CR fluids. Cyan dotted and black solid lines represent, respectively, the Pad\'{e} approximation (\ref{Pade approximation dP}) applied to the fourth-order asymptotic solutions for the Oldroyd-B and FENE-CR fluids.

First, we observe that the fourth-order asymptotic solutions (gray solid and green dashed lines) cannot accurately capture the pressure drop except for very low values of $De$, consistent with results of~\citet{housiadas2023_2D}, indicating that the asymptotic series has a very small radius of convergence. Nevertheless,
when using the Pad\'{e} approximation to accelerate the convergence of the asymptotic series, we find that our analytical predictions for the pressure drop are in excellent agreement with numerical simulations for both Oldroyd-B and FENE-CR fluids. For example, even for $L^2\epsilon^2=0.1$, where the Pad\'{e} approximation slightly overpredicts the pressure drop of the FENE-CR fluid, the relative error is approximately 5 \% for up to $De = 0.5$.

Second, it is evident that, at low Deborah numbers, the dimensionless pressure drop of both Oldroyd-B and FENE-CR fluids monotonically decreases
with $De$, similar to Giesekus and FENE-P fluids~\citep{housiadas2023_2D}. Furthermore, as expected, for $L^2\epsilon^2=10$ and $L^2\epsilon^2=5$, the pressure drop behavior of both fluids is almost indistinguishable.
However, when the finite extensibility becomes more apparent, i.e., as $L^2\epsilon^2$ decreases, the FENE-CR model predicts a higher dimensionless pressure drop than the Oldroyd-B model, as shown in figure~\ref{Fig:2}($d$).

\begin{figure}
 \centerline{\includegraphics[scale=1]{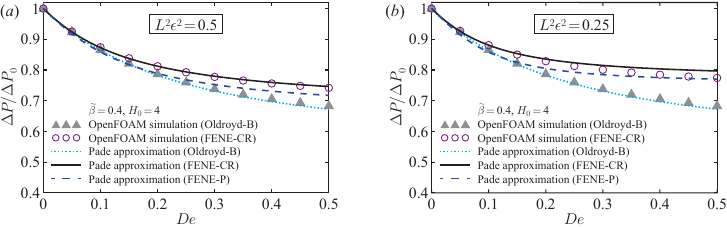}}
\caption{Comparison of non-dimensional pressure drop at low Deborah numbers for the Oldroyd-B, FENE-CR, and FENE-P fluids in a contracting channel. ($a,b$) Scaled pressure drop $\Delta P/\Delta P_0$ as a function of $De=\lambda q/(2\ell h_{\ell})$ for ($a$) $L^2\epsilon^2=0.5$ and ($b$) $L^2\epsilon^2=0.25$. Gray triangles and purple circles represent the OpenFOAM simulation results for the Oldroyd-B and FENE-CR fluids, respectively. Cyan dotted, solid black, and dashed blue lines represent the Pad\'{e} approximation (\ref{Pade approximation dP}) applied to the fourth-order asymptotic solutions for the Oldroyd-B, FENE-CR, and FENE-P fluids. All calculations were performed using $H_0=4$ and $\tilde\beta=0.4$.}\label{Fig:3_New}
\end{figure}

It is of particular interest to compare and contrast our predictions for the pressure drop of the FENE-CR fluid with recent low-$De$ results of~\citet{housiadas2023_2D} for the FENE-P fluid. Such a comparison of the non-dimensional pressure drop is shown in figure~\ref{Fig:3_New} for Oldroyd-B, FENE-CR, and FENE-P fluids in a contracting channel for $L^2\epsilon^2=0.5$ and $0.25$.
Blue dashed lines represent the Pad\'{e} approximation (\ref{Pade approximation dP}) applied to the fourth-order asymptotic solutions obtained from~\citet{housiadas2023_2D} for the FENE-P fluid, when accounting for the differences in characteristic scales. 
Similar to the Oldroyd-B and FENE-CR fluids, the dimensionless pressure drop of the FENE-P fluid monotonically decreases with $De$ at low Deborah numbers.
Furthermore, as expected, the FENE-P fluid shows a higher pressure drop than the Oldroyd-B fluid due to the effects of finite extensibility. However, due to the shear-thinning effects, the resulting pressure drop of the FENE-P fluid is lower than that of the FENE-CR fluid.

Although our low-$De$ analysis using the Pad\'{e} approximation predicts well the pressure drop at low Deborah numbers, it cannot accurately capture the pressure drop behavior at high Deborah numbers. To this end, in the next subsections, we employ numerical simulations and the low-$\tilde \beta$ lubrication analysis.


\subsection{Pressure drop and elastic stresses at high Deborah numbers}\label{dP at high De}

In this subsection, we study and contrast the elastic stresses and pressure drop of the Oldroyd-B and FENE-CR fluids across the contraction at high Deborah numbers. Specifically, we first consider high Deborah numbers up to $De= 4$ using the OpenFOAM simulations and validate the predictions of our low-$\tilde \beta$ lubrication analysis. Then, we employ the low-$\tilde \beta$ lubrication analysis to study the behavior of the elastic stresses and pressure drop at sufficiently high Deborah numbers up to $De=20$.

\begin{figure}
 \centerline{\includegraphics[scale=1]{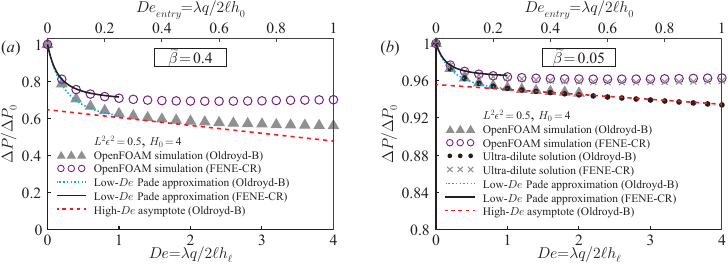}}
\caption{Non-dimensional pressure drop at high Deborah numbers for the Oldroyd-B and FENE-CR fluids in a contracting channel. ($a,b$) Scaled pressure drop $\Delta P/\Delta P_0$ as a function of $De=\lambda q/(2\ell h_{\ell})$ (or $De_{\it entry}=\lambda q/(2\ell h_{0})$) for ($a$) $\tilde\beta=0.4$ and ($b$) $\tilde\beta=0.05$. Gray triangles and purple circles represent the OpenFOAM simulation results for the Oldroyd-B and FENE-CR fluids. Black dots and gray crosses in 
 ($b$) represent the results of the low-$\tilde \beta$ lubrication analysis for the Oldroyd-B and FENE-CR fluids. Cyan dotted and solid black lines represent the low-$De$ Pad\'{e} approximation (\ref{Pade approximation dP}) for the Oldroyd-B and FENE-CR fluids. Red dashed lines represent the high-$De$ asymptotic solution (\ref{Pressure drop OB high-De}) for the Oldroyd-B fluid. All calculations were performed using $H_0=4$ and $L^2\epsilon^2=0.5$.}
\label{Fig:3}
\end{figure}

First, in figure~\ref{Fig:3}($a,b$) we present the scaled pressure drop $\Delta P/\Delta P_0$ of the Oldroyd-B and FENE-CR fluids in the contraction as a function of $De=\lambda q/(2\ell h_{\ell})$ for ($a$) $\tilde\beta=0.4$ and ($b$) $\tilde\beta=0.05$, with $L^2\epsilon^2 = 0.5$. Gray triangles and purple circles represent, respectively, the OpenFOAM simulation results for Oldroyd-B and FENE-CR fluids. Black dots and gray crosses represent, respectively, the results of the low-$\tilde\beta$ lubrication analysis for the Oldroyd-B and FENE-CR fluids. Cyan dotted and solid black lines represent the low-$De$ Pad\'{e} approximation (\ref{Pade approximation dP}) for the Oldroyd-B and FENE-CR fluids. Red dashed lines represent the high-$De$ asymptotic solution (\ref{Pressure drop OB high-De}) for the Oldroyd-B fluid in the ultra-dilute
limit. As both the Deborah number $De=\lambda q/(2\ell h_{\ell})$ based on the exit height and the Deborah number $De_{\it entry}=\lambda q/(2\ell h_{0})$ based on the entry height are used in the literature, we present our results both as a function of $De$ and $De_{\it entry}$.

Consistent with the previous studies~\citep{BoykoHinchStone2023,HinchBoykoStone2023}, the pressure drop of the Oldroyd-B fluid monotonically decreases with $De$ and scales linearly with $De$ at high Deborah
numbers for $\tilde\beta=0.05$, corresponding to the ultra-dilute limit, as represented by the red dashed line in figure~\ref{Fig:3}($b$). Furthermore, there is excellent agreement between the predictions of the low-$\tilde \beta$ lubrication analysis with $\tilde\beta=0.05$ and the OpenFOAM simulations. In particular, for the Oldroyd-B fluid, the relative error at $De =2$ is 0.2 \%.
Nevertheless, as expected, for $\tilde\beta=0.4$ (figure~\ref{Fig:3}($a$)), the high-$De$ asymptotic solution (\ref{Pressure drop OB high-De}) for the Oldroyd-B fluid in the ultra-dilute limit does not accurately capture the slope of the OpenFOAM simulations due to the deviations in the flow velocity from the parabolic profile when $\tilde\beta\nll1$. 

In contrast to a monotonic pressure drop reduction with $De$ observed for the Oldroyd-B fluid, the pressure drop of the FENE-CR fluid levels off to a plateau at high Deborah numbers for both $\tilde\beta=0.4$ and 0.05, with a slight increase for $De\gtrsim3$. 
Understanding this non-monotonic pressure drop variation for the FENE-CR fluid necessitates analyzing higher Deborah numbers.
We note that the presented OpenFOAM simulations for the FENE-CR fluid are in the range of $0\leq De\leq 4$. Performing simulations at higher Deborah numbers requires a longer downstream (exit) section to allow the elastic stresses to reach their fully relaxed values~\citep[see, e.g.,][]{debbaut1988numerical,keiller1993spatial,alves2003benchmark,BoykoHinchStone2023}, thus significantly increasing the computational time. Furthermore, above a certain high $De$, we expect our OpenFOAM simulations to suffer from accuracy and convergence difficulties associated with the high-Weissenberg-number problem~\citep{owens2002computational,alves2021numerical}. Indeed, for the Oldroyd-B fluid with $\tilde\beta=0.05$, we cannot obtain reliable results above $De \approx 2$.

Therefore, instead of carrying out computationally expensive simulations, we use the low-$\tilde \beta$ lubrication analysis considering the ultra-dilute limit, which is considerably faster and allows us to access the behavior of the elastic stresses and pressure drop at arbitrary values of $De$. Such an approach is strongly supported by the excellent agreement between the pressure drop predictions of the low-$\tilde\beta$ lubrication analysis and the OpenFOAM simulation results, as shown in figure~\ref{Fig:3}($b$). Specifically, for the FENE-CR fluid, we find that a relative error is approximately 0.3 \% for up to $De = 4$.

\begin{figure}
 \centerline{\includegraphics[scale=1]{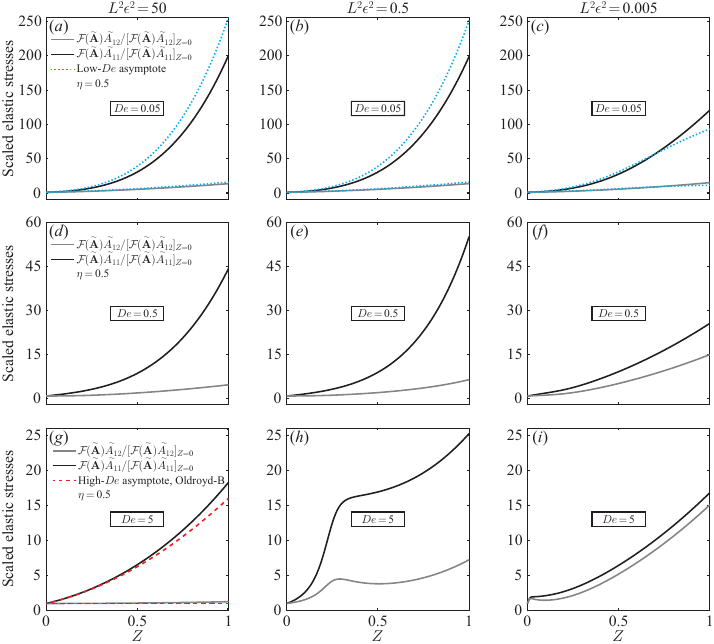}}
\caption{The streamwise variation of elastic stresses of the FENE-CR fluid on $\eta=0.5$ in a contracting channel in the ultra-dilute limit.
($a$--$i$) Elastic normal and shear stresses $\mathcal{F}(\tilde{\mathsfbi{A}})\tilde{A}_{11}$ and $\mathcal{F}(\tilde{\mathsfbi{A}})\tilde{A}_{12}$, scaled by their entry values, as a function of $Z$ for different values of $De$ and $L^2\epsilon^2$.
Solid lines represent the results of the low-$\tilde \beta$ lubrication analysis. Cyan dotted lines in ($a$--$c$) represent the low-$De$ asymptotic solutions for the FENE-CR fluid. Red dashed lines in ($g$) represent the high-$De$ asymptotic solutions for the Oldroyd-B fluid. All calculations were performed using $H_{0}=4$.}\label{Fig:4}
\end{figure}

Before investigating the pressure drop behavior at higher Deborah numbers, it is of particular interest to elucidate the spatial variation of the elastic stresses.
We present in figure~\ref{Fig:4} the streamwise variation of the elastic normal and shear stresses of the FENE-CR fluid, scaled by their entry values, on $\eta=0.5$ in a contracting channel in the ultra-dilute limit for different values of $De$ and $L^2\epsilon^2$.
As expected, for $L^2\epsilon^2=50$, we recover the Oldroyd-B behavior previously studied by~\citet{HinchBoykoStone2023} and \citet{BoykoHinchStone2023}. Specifically, we find that, at low Deborah numbers ($De=0.05$, figure~\ref{Fig:4}($a$)), the elastic shear and normal stresses increase by a factor of $H_0^2=16$ and $H_0^4=256$, respectively, by the end of contraction. In contrast, at high Deborah numbers ($De=5$, figure~\ref{Fig:4}($g$)), the elastic shear stress $\mathcal{F}(\tilde{\mathsfbi{A}})\tilde{A}_{12}$ preserves its entry value and the elastic normal stress $\mathcal{F}(\tilde{\mathsfbi{A}})\tilde{A}_{11}$ increases by a factor of $H_0^2=16$.

It is evident from figure~\ref{Fig:4}($a$--$c$) that at $De=0.05$, the elastic shear stress weakly depends on the finite extensibility parameter $L^2\epsilon^2$, where the magnitude of the elastic normal stress decreases as $L^2\epsilon^2$ is reduced from 50 to 0.005. At higher Deborah numbers, $De=0.5$ and $De=5$,  we observe a trade-off between the axial component of the conformation tensor $\tilde{A}_{11}$ and the finite extensibility, incorporated by the nonlinear spring function $\mathcal{F}(\tilde{\mathsfbi{A}})=(1-\tilde{A}_{11}/(\epsilon^2 L^2))^{-1}$.
On the one hand, when $L^2\epsilon^2$ is large (the Oldroyd-B limit), the dumbbell extension, as measured by $\mathrm{tr}\tilde{\mathsfbi{A}}\approx\tilde{A}_{11}$, is large and $\mathcal{F}(\tilde{\mathsfbi{A}})\approx 1$. On the other hand, when $L^2\epsilon^2$ is small, the dumbbell extension $\mathrm{tr}\tilde{\mathsfbi{A}}\approx\tilde{A}_{11}$ is also small but $\mathcal{F}(\tilde{\mathsfbi{A}})$ can be large. Therefore, as shown in figures~\ref{Fig:4}($d$--$f$) and~\ref{Fig:4}($g$--$i$), for a sufficient large $De$, the maximum value of elastic normal stress $\mathcal{F}(\tilde{\mathsfbi{A}})\tilde{A}_{11}$, achieved at the end of contraction, may exhibit a non-monotonic variation with $L^2\epsilon^2$ (see also figure~\ref{Fig:7}($b$)). For example, when $De=5$, the maximum value of $\mathcal{F}(\tilde{\mathsfbi{A}})\tilde{A}_{11}$ for  $L^2\epsilon^2=0.5$ is greater than the corresponding values for $L^2\epsilon^2=50$ and $L^2\epsilon^2=0.005$. 
Furthermore, for $De=5$, in contrast to the Oldroyd-B fluid where $\mathcal{F}(\tilde{\mathsfbi{A}})\tilde{A}_{12}$ maintains its entry value, when the finite extensibility is significant, i.e., $L^2\epsilon^2=0.5$ and $L^2\epsilon^2=0.005$, we observe a non-monotonic increase of elastic shear stress with axial position $Z$.

\begin{figure}
 \centerline{\includegraphics[scale=1]{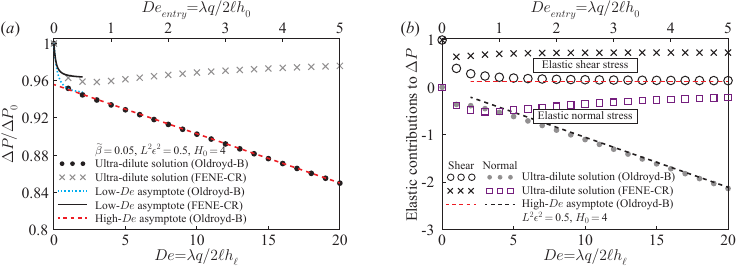}}
\caption{($a$) Scaled pressure drop $\Delta P/\Delta P_0$ as a function of $De=\lambda q/(2\ell h_{\ell})$ for $\tilde\beta=0.05$. Black dots and gray crosses represent the results of the low-$\tilde \beta$ lubrication analysis for the Oldroyd-B and FENE-CR fluids.  Cyan dotted and solid black lines represent the low-$De$ Pad\'{e} approximation (\ref{Pade approximation dP}) for the Oldroyd-B and FENE-CR fluids. The red dashed line represents the high-$De$ asymptotic solution (\ref{Pressure drop OB high-De}) for the Oldroyd-B fluid.
 ($b$) Elastic contributions to the non-dimensional pressure drop, scaled by $\tilde\beta$, as a function of $De=\lambda q/(2\ell h_{\ell})$ in the ultra-dilute limit. Black circles and gray dots represent ultra-dilute predictions of the Oldroyd-B fluid for elastic shear and normal stress contributions. Black crosses and purple squares represent ultra-dilute predictions of the FENE-CR fluid for elastic shear and normal stress contributions. Red and black dashed lines represent the high-$De$ asymptotic solution of the Oldroyd-B fluid for elastic shear and normal stress contributions. All calculations were performed using $H_0=4$ and $L^2\epsilon^2=0.5$.}
\label{Fig:5}
\end{figure}

Next, we analyze the pressure drop variation at significantly higher Deborah numbers using our low-$\tilde\beta$ lubrication analysis.
We present in figure~\ref{Fig:5}($a$) the scaled pressure drop $\Delta P/\Delta P_0$ of the Oldroyd-B and FENE-CR fluids in the contraction as a function of $De=\lambda q/(2\ell h_{\ell})$ for $L^2\epsilon^2 = 0.5$ and $\tilde\beta=0.05$, corresponding to the ultra-dilute limit. 
Black dots represent the results of the low-$\tilde\beta$ lubrication analysis, the cyan dotted line represents the low-$De$ Pad\'{e} approximation (\ref{Pade approximation dP}), and the red dashed line represents the high-$De$ asymptotic solution (\ref{Pressure drop OB high-De}) for the Oldroyd-B fluid. We observe excellent agreement between our low- and high-$De$ asymptotic solutions and the low-$\tilde\beta$ lubrication results. Moreover, somewhat surprisingly, from figure~\ref{Fig:5}($a$) it follows that the low-$De$ Pad\'{e} approximation (\ref{Pade approximation dP}) captures fairly well the pressure drop reduction with $De$ for up to $De=2$ ($De_{\it entry}=0.5$) for both Oldroyd-B and FENE-CR fluids. More importantly, unlike a linear pressure drop reduction of the Oldroyd-B fluid at high Deborah numbers, the pressure drop of the FENE-CR fluid (gray crosses) exhibits a non-monotonic variation, first decreasing with $De$, attaining a local minimum at $De \approx 2.8$, and then increasing with $De$. Such a non-monotonic variation in the pressure drop is consistent with the previous numerical studies on the flow of the FENE-P fluid in 2-D abruptly contracting geometries~\citep{zografos2022viscoelastic}. Nevertheless, the non-dimensional pressure drop for the FENE-CR fluid in the ultra-dilute limit is lower than the corresponding Newtonian pressure drop, i.e., $\Delta P/\Delta P_0<1$, even for very high Deborah numbers.  

To probe deeper into the source of the non-monotonic variation of the pressure drop for the FENE-CR fluid, we present in figure~\ref{Fig:5}($b$) the elastic contributions to the non-dimensional pressure drop, scaled by $\tilde\beta$, as a function of $De=\lambda q/(2\ell h_{\ell})$ in the ultra-dilute limit.
Black circles and gray dots represent the elastic shear and normal stress
contributions obtained from the low-$\tilde\beta$ lubrication analysis for the Oldroyd-B fluid. Black crosses and purple squares represent the elastic shear and normal stress
contributions obtained from the low-$\tilde\beta$ lubrication analysis for the FENE-CR fluid. As expected, for the Oldroyd-B fluid, there is excellent agreement between our low-$\tilde \beta$ lubrication results and the high-$De$ asymptotic solution (\ref{Pressure drop OB high-De}), represented by red and black dashed lines.

In contrast to the Oldroyd-B fluid, where the elastic normal stress contribution decreases with $De$ and scales linearly with $De$ at high Deborah numbers, for the FENE-CR fluid we observe a non-monotonic variation. In particular, the elastic normal stress contribution of the FENE-CR fluid first decreases, attains a minimum at $De \approx 3.2$, and then increases with $De$. Such an increase is associated with the dissipative effect of the finite extensibility. Despite this increase, figure~\ref{Fig:5}($b$) clearly shows that the elastic normal stress contribution of the FENE-CR fluid is negative at $De=20$, leading to a reduction in the pressure drop, similar to the Oldroyd-B fluid. However, we find that, at $De \approx 118$, the elastic normal stress contribution of the FENE-CR fluid becomes positive and then increases with $De$. For $\tilde\beta=0.05$, we have confirmed that up to $De=1000$, this positive elastic normal stress contribution is too weak since it scales with $\tilde\beta$, and thus cannot lead to the pressure drop enhancement above the Newtonian value $\Delta P_0$. Note that we have assumed steady flows, so further investigation is necessary to determine if there might be flow instabilities at these high Deborah numbers. Nevertheless, as pointed out by~\citet{HinchBoykoStone2023}, under the lubrication approximation, the hoop stress is neglected, so purely elastic instability cannot arise due to curved streamlines.
\begin{figure}
 \centerline{\includegraphics[scale=1]{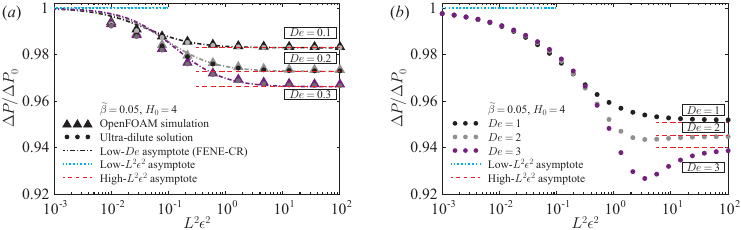}}
\caption{The influence of the finite extensibility on the non-dimensional pressure drop of the FENE-CR fluid in a contracting channel. ($a,b$) Scaled pressure drop $\Delta P/\Delta P_0$ as a function of the finite extensibility parameter $L^2\epsilon^2$ for ($a$) low- and ($b$) high Deborah numbers. Triangles in ($a$) represent the OpenFOAM simulation results. Dots represent the results obtained from the low-$\tilde{\beta}$ lubrication analysis. Dashed-dotted lines represent the low-$De$~Pad\'{e} approximation (\ref{Pade approximation dP}) applied up to the fourth-order asymptotic solution. Cyan dotted lines represent the low-$L^2\epsilon^2$ asymptotic solution, corresponding to the Newtonian limit. Red dashed lines represent the high-$L^2\epsilon^2$ asymptotic solution, corresponding to the Oldroyd-B limit. All calculations were performed using $H_0=4$ and $\tilde\beta=0.05$.}\label{Fig:6}
\end{figure}

The elastic shear stress contribution of the FENE-CR fluid also exhibits a non-monotonic variation with the Deborah number. It first decreases, attains a minimum at $De \approx 1.2$, and then approaches a plateau at high Deborah numbers. 
Such a non-monotonic variation of the elastic normal and shear stress contributions rationalizes the non-monotonic pressure drop behavior, shown in figure~\ref{Fig:5}($a$).
Similar to the Oldroyd-B fluid, the elastic shear stress contribution of the FENE-CR fluid is independent of $De$ at high Deborah numbers, but with a constant value higher than for the Oldroyd-B fluid, due to the dissipative effect of the finite extensibility. This higher value of elastic shear stress contribution leads to an even greater increase in the pressure drop of the FENE-CR fluid compared with the Oldroyd-B fluid.

\subsection{Assessing the effect of the finite extensibility on the pressure drop}

In the previous subsections, we analyzed the pressure drop variation with the Deborah number $De$ and the viscosity ratio $\tilde{\beta}$, mainly considering the finite extensibility parameter $L^2\epsilon^2=0.5$. In this subsection, we study how the finite extensibility parameter $L^2\epsilon^2$ impacts the pressure drop.

First, in figure~\ref{Fig:6}($a,b$) we present the variation of the scaled pressure drop $\Delta P/\Delta P_0$ as a function of $L^2\epsilon^2$ for the FENE-CR fluid in a contracting channel for ($a$) low- and ($b$) high Deborah numbers, with $\tilde\beta=0.05$. Triangles and dots represent, respectively, the results of the OpenFOAM simulations and low-$\tilde{\beta}$ lubrication analysis. Dashed-dotted lines represent the low-$De$~Pad\'{e} approximation (\ref{Pade approximation dP}) applied up to the fourth-order asymptotic solution. Cyan dotted and red dashed lines represent the low- and high-$L^2\epsilon^2$ asymptotic solutions, corresponding to the Newtonian and Oldroyd-B limits, respectively.

At low Deborah numbers, it is evident from figure~\ref{Fig:6}($a$) that, the pressure drop monotonically decreases with increasing $L^2\epsilon^2$. Clearly, there is excellent agreement between our low-$De$ asymptotic solutions based on the Pad\'{e} approximation, the OpenFOAM simulation results, and the predictions of the low-$\tilde{\beta}$ lubrication analysis.
Consistent with the low-$De$~Pad\'{e} approximation (\ref{Pade approximation dP}), for small values of $L^2\epsilon^2$, the pressure drop becomes independent of $De$, approaching the Newtonian limit for all values of $De$, represented by cyan dotted lines. As expected, for large values of $L^2\epsilon^2$, the pressure drop approaches the Oldroyd-B limit, represented by red dashed lines.

Next, we consider the variation in pressure drop with $L^2\epsilon^2$ at high Deborah numbers, as shown in figure~\ref{Fig:6}($b$). 
At high Deborah numbers, the pressure drop shows Newtonian and Oldroyd-B asymptotic behavior for $L\epsilon \ll 1$ and $L\epsilon \gg 1$, similar to the low-$De$ limit.
However, in contrast to low Deborah numbers, at high Deborah numbers $De=2$ and 3, pressure drop exhibits a strong non-monotonic
behavior with $L^2\epsilon^2$. Specifically, we observe that the pressure drop first decreases and then increases with  $L^2\epsilon^2$ approaching the Oldroyd-B limit, with the transition occurring at $L^2\epsilon^2 =O(1)$. 
\begin{figure}
 \centerline{\includegraphics[scale=1]{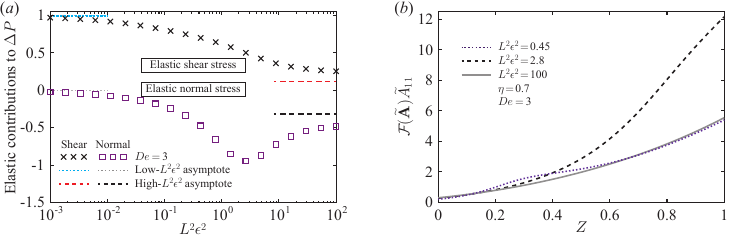}}
\caption{($a$) Elastic contributions to the non-dimensional pressure drop of the FENE-CR fluid, scaled by $\tilde\beta$, as a function of the finite extensibility parameter $L^2\epsilon^2$ for $De=3$ in the ultra-dilute limit. Black crosses and purple squares represent the elastic shear and normal stress contributions obtained from the low-$\tilde{\beta}$ lubrication analysis. Cyan and gray dotted lines represent the low-$L^2\epsilon^2$ asymptotic solution for the elastic shear and normal stress contributions, corresponding to the Newtonian limit. Red and black dashed lines represent the high-$L^2\epsilon^2$ asymptotic solution (\ref{Pressure drop OB high-De}) for the elastic shear and normal stress contributions, corresponding to the Oldroyd-B limit at high $De$. ($b$) Elastic normal stress $\mathcal{F}(\tilde{\mathsfbi{A}})\tilde{A}_{11}(Z,\eta=0.7)$ as a function of $Z$ for $De=3$ and $L^2\epsilon^2=0.45$ (dotted line), $L^2\epsilon^2=2.8$ (dashed line), and $L^2\epsilon^2=100$ (solid line).
All calculations were performed using $H_0=4$.}\label{Fig:7}
\end{figure}

To provide further insight into the pressure drop dependence on the finite extensibility $L^2\epsilon^2$ for a given $De$, we study the relative importance of elastic contributions to the pressure drop.
The elastic contributions to the non-dimensional pressure drop across the contraction, scaled by $\tilde{\beta}$, as a function of $L^2\epsilon^2$ are shown in figure~\ref{Fig:7}($a$) for $De=3$.
 Black crosses and purple squares represent the elastic shear and normal stress contributions obtained from the low-$\tilde{\beta}$ lubrication analysis. Red and black dashed lines represent the high-$L^2\epsilon^2$ asymptotic solution (\ref{Pressure drop OB high-De}) for the elastic shear and normal stress contributions, corresponding to the Oldroyd-B limit at high $De$. 

For small values of $L^2\epsilon^2$, the elastic normal stress contribution to the pressure drop approaches zero, while the elastic shear stress contribution approaches an order-one Newtonian value. We rationalize this behavior by noting 
 from  
(\ref{Non-dim Warner spring curve}) and (\ref{A_11 BC entrance})--(\ref{A_12 BC entrance}) that at the beginning of the
contraction, in the low-$L^2\epsilon^2$ limit, we have
\begin{equation}
\tilde{A}_{11}=L^2\epsilon^2 -\frac{H_0^2 L^3\epsilon^3}{3 \sqrt{2} De \eta }+O(L^4\epsilon^4), \quad \tilde{A}_{12}=-\frac{L\epsilon}{\sqrt{2}}+\frac{H_0^2 L^2\epsilon^2 }{12 De\eta }+O(L^3\epsilon^3) \quad \text{for} \; L\epsilon \ll 1,  \label{A ent 1} 
\end{equation}
\begin{equation}
\mathcal{F}(\tilde{\mathsfbi{A}})\tilde{A}_{11}=\frac{3 \sqrt{2} L\epsilon De  }{H_0^2}\eta+O(L^2\epsilon^2), \quad \mathcal{F}(\tilde{\mathsfbi{A}})\tilde{A}_{12}=-\frac{3 De  }{H_0^2}\eta \quad \text{for} \; L\epsilon \ll 1.  \label{A ent 2} 
\end{equation}
This result is valid for all $De$. Therefore, for $L\epsilon \ll 1$, the elastic normal stress $\mathcal{F}(\tilde{\mathsfbi{A}})\tilde{A}_{11}$ scales as $O(L\epsilon De )$ and the elastic shear stress $\mathcal{F}(\tilde{\mathsfbi{A}})\tilde{A}_{12}$ scales as $O(De)$. Using (\ref{dP non-uniform by parts FENE-CR CC}), the latter scaling arguments imply that the elastic normal stress contribution to the pressure drop scales as $O(L\epsilon )$ and thus is negligible for all $De$. On the other hand, the elastic shear stress has a Newtonian contribution, which is independent of $De$, as shown in figure~\ref{Fig:7}($a$).

Furthermore, we observe that, while the elastic shear stress contribution monotonically decreases with increasing $L^2\epsilon^2$, the elastic normal stress contribution exhibits a non-monotonic variation with $L^2\epsilon^2$. 
Thus, the non-monotonic behavior of the pressure drop, shown in figure~\ref{Fig:6}($b$) for $De=2$ and 3 at $L^2\epsilon^2 =O(1)$, arises due to the elastic normal stress contribution. 
Such a non-monotonic variation with $L^2\epsilon^2$ for a given $De$ can be attributed to the trade-off between the axial component of the conformation tensor $\tilde{A}_{11}$ and the finite extensibility $L^2 \epsilon^2$ through $\mathcal{F}(\tilde{\mathsfbi{A}})=(1-\tilde{A}_{11}/(\epsilon^2 L^2))^{-1}$,
as discussed in $\mathsection$~\ref{dP at high De}. For example, as shown in figure~\ref{Fig:7}($b$), for a given $De$, the elastic normal stress $\mathcal{F}(\tilde{\mathsfbi{A}})\tilde{A}_{11}$ can exhibit similar spatial variations for small (dotted line) and large (solid line) values of $L^2\epsilon^2$, rationalizing the non-monotonic behavior of elastic normal stress contribution to the pressure drop. 


\section{Concluding remarks}\label{CR}

In this work, we studied the flow of a FENE-CR fluid in slowly varying contracting channels at low and high Deborah numbers. 
Employing the low-Deborah-number lubrication analysis, we provided analytical expressions for the non-dimensional pressure drop for the FENE-CR fluid up to $O(De^4)$ and applied the Pad\'{e} approximation to improve the convergence of the asymptotic series. To understand the pressure drop behavior of the FENE-CR fluid at high Deborah numbers, we considered the ultra-dilute limit of small polymer concentration and exploited the one-way coupling between the parabolic velocity and elastic stresses to calculate the pressure drop for arbitrary values of $De$.
We further compared and contrasted the predictions of the FENE-CR model to the recent results of~\citet{BoykoHinchStone2023} and~\citet{HinchBoykoStone2023} for the Oldroyd-B model as well as to the low-$De$ results of~\citet{housiadas2023_2D} for the FENE-P model.
We validated our theoretical results for the dimensionless pressure drop in a contracting channel with 2-D finite-volume numerical simulations for both Oldroyd-B and FENE-CR fluid and found excellent agreement.

At low Deborah numbers, the pressure drop of the FENE-CR fluid monotonically decreases with $De$, as shown in figure~\ref{Fig:2}, similar to the predictions of the Oldroyd-B and FENE-P fluids. However, at high Deborah numbers, unlike a linear pressure drop reduction of the Oldroyd-B fluid, the pressure drop of the FENE-CR fluid exhibits a non-monotonic variation, first decreasing and then increasing with $De$. Note that the pressure drop for the FENE-CR fluid remains lower than the corresponding Newtonian pressure drop even for very high Deborah numbers, as shown in figure~\ref{Fig:5}($a$). We identified two causes for such pressure drop variation of the FENE-CR fluid (see figure~\ref{Fig:5}($b$)). The first cause is the elastic normal stress contribution to the pressure drop, which becomes less negative as $De$ increases at high Deborah numbers due to the dissipative effect of the finite extensibility. The second cause is the contribution of elastic shear stresses, which is higher compared to the Oldroyd-B fluid, again owing to the dissipative effect of the finite extensibility. 

In general, the pressure drop of the FENE-CR fluid increases compared with the Oldroyd-B fluid as the finite extensibility becomes more apparent (when $L^2\epsilon^2$ decreases). Nevertheless, for very small values of $L^2\epsilon^2$, the pressure drop of FENE-CR fluid becomes independent of $De$ and approaches the Newtonian value. Specifically, when $L^2\epsilon^2 \ll1$, the elastic normal stress contribution vanishes while the elastic shear stress contribution shows a Newtonian behavior for all $De$ (see figure~\ref{Fig:6} and figure~\ref{Fig:7}($a$)).

Our theoretical framework, based on lubrication theory and the ultra-dilute limit, allows us to study the behavior of the elastic stresses and pressure drop of a FENE-CR fluid at sufficiently high Deborah numbers. We emphasize that we are currently unable to achieve these high values of the Deborah number using finite-volume or finite-element simulations.
We, therefore,  believe that our theoretical results for the FENE-CR fluid in the ultra-dilute limit, valid at all $De$, are of fundamental interest and can be helpful for simulation validation and enhancing our understanding of viscoelastic channel flows.

The theoretical predictions of the non-monotonic pressure drop behavior of the FENE-CR fluid in a contraction are consistent with the previous numerical studies on contraction geometries~\citep[see, e.g.,][]{nystrom2012numerical,zografos2022viscoelastic}. However, these predictions are in contrast with the experimental results showing a nonlinear increase in the pressure drop with $De$ above the Newtonian pressure drop value for the flow of a Boger fluid through abrupt axisymmetric contraction and contraction--expansion  geometries~\citep{rothstein1999extensional,rothstein2001axisymmetric,nigen2002viscoelastic,sousa2009three}. Our results with the FENE-CR model that incorporates the feature of finite extensibility cannot resolve this contradiction. 
Thus, as a future research direction, it is interesting to study more
complex elastic dumbbell models that account for additional microscopic features of realistic polymer chains, such as the conformation-dependent friction coefficient and the conformation-dependent non-affine deformation~\citep{phan1984study,BoykoStone2024Perspective}, and to elucidate their effect on the pressure drop. 

Finally, we note that, in this work, we have focused on studying the pressure drop across the contraction region. However, numerical simulations and experimental set-ups include a long downstream (exit) section to allow the stresses to reach their fully relaxed values~\citep{keiller1993spatial,rothstein2001axisymmetric,BoykoHinchStone2023}. Therefore, one interesting extension of the present work is to study the spatial relaxation of elastic stresses, velocity, and pressure of viscoelastic fluids in the exit channel using the FENE-CR model and more complex constitutive equations.

\backsection[Supplementary material]{Supplementary material includes the \textsc{Mathematica} file containing the explicit expressions for the velocity, conformation tensor components, and the pressure drop in the low-Deborah-number limit up to $O(De^4)$.}

\backsection[Funding]{B.M.\ and E.B.\ acknowledge the support by grant no. 2022688 from the US-Israel Binational Science Foundation (BSF) and grant no. 1942/23 from the Israel Science Foundation (ISF). H.A.S.\ acknowledges the support from grant no. CBET-2246791 from the United States National Science Foundation (NSF). E.B.\ acknowledges the support from the Israeli Council for Higher Education Yigal Alon Fellowship.}

\backsection[Declaration of interests]{The authors report no conflict of interest.}

\backsection[Author ORCIDs]{
\\Bimalendu Mahapatra https://orcid.org/0000-0003-0845-9775;\\
Tachin Ruangkriengsin https://orcid.org/0000-0002-5422-4822;\\
Howard A. Stone https://orcid.org/0000-0002-9670-0639;\\
Evgeniy Boyko https://orcid.org/0000-0002-9202-5154.}


\appendix

\section{A fully developed flow of a FENE-CR fluid in a straight channel }\label{App: straight channel}

Consider a steady and fully developed flow of a FENE-CR fluid in a straight and long channel of non-dimensional height $2H_0$. 
Under the assumption of a fully developed flow, we have $ U_y\equiv 0$, so that the governing equations (\ref{Non-dim Gov.Eq FCR}) simplify to
\begin{subequations}\label{Non-dim Gov.Eq FCR straight channel}
\begin{equation}\label{Non-dim.z-momentum FCR straight channel}
\frac{\mathrm{d} P}{\mathrm{d} Z}=(1-\tilde\beta)\frac{\mathrm{d}^2 U_z}{\mathrm{d} Y^2}+\frac{\tilde\beta}{De} \frac{\mathrm{d}(\mathcal{F}(\tilde{\mathsfbi{A}})\tilde A_{yz})}{\mathrm{d} Y},
\end{equation}
\begin{equation}\label{Non-dim Azz FCR straight channel}
2\frac{\mathrm{d} U_z}{\mathrm{d} Y}\tilde A_{yz}=\frac{\mathcal{F}(\tilde{\mathsfbi{A}})}{De}\tilde{A}_{zz},
\end{equation}
\begin{equation}\label{Non-dim Ayz FCR straight channel} 
\frac{\partial U_z}{\partial Y}\tilde A_{yy}=\frac{\mathcal{F}(\tilde{\mathsfbi{A}})}{De}\tilde{A}_{yz},
\end{equation}
\begin{equation}\label{Non-dim Ayy FCR straight channel}
\tilde{A}_{yy}=1.
\end{equation}
\end{subequations}
Substituting (\ref{Non-dim Ayz FCR straight channel}) into (\ref{Non-dim.z-momentum FCR straight channel}) yields 
\begin{equation}\label{straight channel z-momentum 2}
\frac{\mathrm{d} P}{\mathrm{d} Z}=\frac{\mathrm{d}^2 U_z}{\mathrm{d} Y^2}.
\end{equation}
Solving for the velocity $U_z$ subject to (\ref{ND BC FCR Velocity}), we obtain a parabolic profile
\begin{equation} \label{Solution Uz straight channel}
    U_{z}(Y)=\frac{3}{2}\frac{H_{0}^2-Y^2}{H_{0}^3}.
\end{equation}
Next, substituting (\ref{Non-dim Ayz FCR straight channel}) into (\ref{Non-dim Azz FCR straight channel}) and using (\ref{Non-dim Ayy FCR straight channel}) and (\ref{Non-dim Warner spring FCR}) leads to the nonlinear algebraic equation for $\tilde A_{zz}$
\begin{equation}
    2De^2\left(1-\frac{\tilde A_{zz}}{L^2\epsilon^2}\right)^2\left(\frac{\partial U_z}{\partial Y}\right)^2=\tilde A_{zz}.\label{Eq. Azz straight channel}
\end{equation}
The corresponding solution of (\ref{Eq. Azz straight channel}) is
\begin{equation}\label{sol. straight channel Azz}
\tilde A_{zz}=L^2\epsilon^2+L^3\epsilon^3\frac{L\epsilon-\sqrt{L^2\epsilon^2+72De^2 Y^2/H^6_0}}{36 De^2 Y^2/H^6_0}.
\end{equation}
Combining (\ref{Non-dim Ayz FCR straight channel}) and (\ref{sol. straight channel Azz}) provides the expression for $\tilde A_{yz}$
\begin{equation}\label{sol. straight channel Ayz}
\tilde A_{yz}=L\epsilon\frac{L\epsilon-\sqrt{L^2\epsilon^2+72De^2 Y^2/H^6_0}}{12De Y/H^3_0}.
\end{equation}
Finally, we note that considering the limit $L^2\epsilon^2 \to \infty$ and using (\ref{sol. straight channel Azz})--(\ref{sol. straight channel Ayz}), we obtain the corresponding expressions for the conformation tensor components of the Oldroyd-B fluid
\begin{equation}
    \tilde{A}_{zz}=\frac{18De^{2}}{H_{0}^{6}}Y^{2},\qquad \tilde{A}_{yz}=-\frac{3De}{H_{0}^{3}}Y ,\qquad \tilde{A}_{yz}=1.\label{sol. straight channel Azz Ayz OB} 
\end{equation}

\section{Low-Deborah-number lubrication analysis: detailed derivation}\label{Details of low De}

We here provide details of the derivation of the analytical expressions for the pressure drop of the FENE-CR fluid in the low-$De$ limit up to $O(De^4)$.

Before proceeding to the asymptotic solution of the pressure drop, we expand $\mathcal{F}(\tilde{\mathsfbi{A}}) \tilde A_{zz}$, $\mathcal{F}(\tilde{\mathsfbi{A}}) \tilde A_{yz}$, and $\mathcal{F}(\tilde{\mathsfbi{A}})(\tilde A_{yy}-1)$ into perturbation series in $De\ll1$. Specifically, using (\ref{Non-dim Warner spring FCR}), (\ref{perturbation_expansion FCR}), (\ref{Leading-order Azz Ayz Ayy FCR}) and noting that $\tilde A_{zz,1}=0$, we obtain
\begin{subequations}\label{Series expansion FCR}
\begin{equation} 
    \mathcal{F}(\tilde{\mathsfbi{A}})\tilde A_{zz}=De^2\tilde A_{zz,2}+De^3\tilde A_{zz,3}+De^4 \left[\tilde A_{zz,4}+\frac{(\tilde A_{zz,2})^2}{L^2 \epsilon ^2}\right]+O(De^5),
  \end{equation} 
\begin{eqnarray}
\mathcal{F}(\tilde{\mathsfbi{A}}) \tilde A_{yz}&=&De\tilde A_{yz,1}+De^2\tilde A_{yz,2}+De^3 \left[\tilde A_{yz,3}+\frac{\tilde A_{yz,1} \tilde A_{zz,2}}{L^2 \epsilon ^2}\right] \nonumber\\
&&+De^4 \left[\tilde A_{yz,4}+\frac{\tilde A_{yz,2}\tilde A_{zz,2}+\tilde A_{yz,1}\tilde A_{zz,3}}{L^2 \epsilon ^2}\right]+O(De^5), 
\end{eqnarray}
\begin{eqnarray}
\mathcal{F}(\tilde{\mathsfbi{A}}) (\tilde A_{yy}-1)&=&De\tilde A_{yy,1}+De^2\tilde A_{yy,2}+De^3 \left[\tilde A_{yy,3}+\frac{\tilde A_{yy,1} \tilde A_{zz,2}}{L^2 \epsilon ^2}\right] \nonumber\\
&&De^4 \left[\tilde A_{yy,4}+\frac{\tilde A_{yy,2}\tilde A_{zz,2}+\tilde A_{yy,1}\tilde A_{zz,3}}{L^2 \epsilon ^2}\right]+O(De^5). 
\end{eqnarray}\end{subequations}

\subsection{Leading-order solution for the pressure drop of a FENE-CR fluid}

Substituting (\ref{perturbation_expansion FCR}) into (\ref{Non-dim Gov.Eq FCR}) and considering the leading order in $De$ and using (\ref{First-order Azz Ayz Ayy FCR}) and (\ref{Series expansion FCR}), we obtain
\refstepcounter{equation}
$$
\frac{\partial U_{z,0}}{\partial Z}+\frac{\partial U_{y,0}}{\partial Y}=0,\qquad \frac{\mathrm{d} P_0}{\mathrm{d} Z}=(1-\tilde\beta)\frac{\partial^2 U_z}{\partial Y^2}+\tilde\beta \frac{\partial \tilde A_{yz,1}}{\partial Y}=\frac{\partial^2 U_{z,0}}{\partial Y^2},\eqno{(\theequation{a,b})}\label{Leading-order Gov.Eq FCR}
$$
subject to the boundary conditions
\refstepcounter{equation}
$$
 U_{z,0}(H(Z),Z)=0, \; U_{y,0}(H(Z),Z)=0, \; \frac{\partial U_{z,0}}{\partial Y}(0,Z)=0, \; \int^{H(Z)}_0 U_{z,0}(Y,Z)\mathrm{d}Y=1. \eqno{(\theequation{a\mbox{$-$}d})}\label{Leading-order BC FCR}
$$As expected, (\ref{Leading-order Gov.Eq FCR}$b$) is the classical momentum equation of the Newtonian fluid with a constant viscosity $\mu_0$. 
The leading-order solutions, previously derived by~\citet{boyko2022pressure}, are given as 
\refstepcounter{equation}
$$
 \mbox{\hspace{-2mm}}U_{z,0}=\frac{3}{2}\frac{H(Z)^2-Y^2}{H(Z)^3}, \; U_{y,0}=\frac{3}{2}\frac{H'(Z)Y(H(Z)^2-Y^2)}{H(Z)^4}, \;  \Delta P_{0}=3\int_{0}^{1}\frac{\mathrm{d}Z}{H(Z)^{3}}, \eqno{(\theequation{a\mbox{$-$}c})}\label{Leading-order sol FCR}
$$where primes indicate derivatives with respect to $Z$.

\subsection{First-order solution for the pressure drop of a FENE-CR fluid}\label{FO correction to PD FCR}

Substituting (\ref{perturbation_expansion FCR}) and (\ref{Series expansion FCR}) into (\ref{Non-dim Gov.Eq FCR}) and considering the first order in $De$, we obtain
\begin{subequations}
\begin{equation} 
\frac{\partial U_{z,1}}{\partial Z}+\frac{\partial U_{y,1}}{\partial Y}=0,\qquad \frac{\mathrm{d} P_1}{\mathrm{d} Z}=(1-\tilde\beta)\frac{\partial^2 U_{z,1}}{\partial Y^2}+\tilde\beta\left( \frac{\partial \tilde A_{zz,2}}{\partial Z}+\frac{\partial \tilde A_{yz,2}}{\partial Y}\right),  \tag{\theequation \emph{a,b}}
\end{equation} 
\begin{equation} 
2\frac{\partial U_{z,0}}{\partial Y}\tilde A_{yz,1}=\tilde A_{zz,2}, \tag{\theequation \emph{c}} 
\end{equation} 
\begin{equation} 
 U_{z,0}\frac{\partial \tilde A_{yz,1}}{\partial Z}+U_{y,0}\frac{\partial \tilde A_{yz,1}}{\partial Y}-\frac{\partial U_{z,0}}{\partial Y}\tilde A_{yy,1}-\frac{\partial U_{z,1}}{\partial Y}=-\tilde A_{yz,2},\tag{\theequation \emph{d}} 
\end{equation} 
\begin{equation} 
U_{z,0}\frac{\partial \tilde A_{yy,1}}{\partial Z}+U_{y,0}\frac{\partial \tilde A_{yy,1}}{\partial Y}-2\frac{\partial U_{y,0}}{\partial Z}\tilde A_{yz,1}-2\frac{\partial U_{ y,0}}{\partial Y}\tilde A_{yy,1}-2\frac{\partial U_{y,1}}{\partial Y}=-\tilde A_{yy,2}.\tag{\theequation \emph{e}} 
\end{equation} \label{First-order New-Gov.Eq FCR} \end{subequations}
These governing equations are supplemented by the boundary conditions 
\refstepcounter{equation}
$$
 U_{z,1}(H(Z),Z)=0, \; U_{y,1}(H(Z),Z)=0, \; \frac{\partial U_{z,1}}{\partial Y}(0,Z)=0, \; \int^{H(Z)}_0 U_{z,1}(Y,Z)\mathrm{d}Y=0. \eqno{(\theequation{a\mbox{$-$}d})}\label{First-order BC FCR}
$$At the first order in $De$, the dimensionless governing equations for the FENE-CR fluid are equivalent to those of the Oldroyd-B fluid. Thus, from (\ref{First-order New-Gov.Eq FCR}), it follows that the expressions for the velocity and pressure drop at $O(De)$ as well as $\tilde A_{zz,2}$, $\tilde A_{yz,2}$, $\tilde A_{yy,2}$ are identical for the FENE-CR and Oldroyd-B fluids, and are given by~\citep{boyko2022pressure}
\begin{subequations}\label{First-order sol FCR}
\begin{gather}
     U_{z,1}\equiv0, \qquad  U_{y,1}\equiv0, \qquad \Delta P_1 = \frac{9}{2}\tilde \beta \left(\frac{1}{H(0)^4}-\frac{1}{H(1)^4} \right), \tag{\theequation \emph{a--c}} \\ 
\tilde A_{zz,2}=\frac{18 Y^2}{H(Z)^6}, \qquad \tilde A_{yz,2}=\frac{18 Y \left(2 Y^2-H(Z)^2\right) H'(Z)}{H(Z)^7}, \tag{\theequation \emph{d,e}}\\ 
\tilde A_{yy,2}=\frac{9}{2}\frac{4\left(-2Y^{2}+H(Z)^{2}\right)^{2}H'(Z)^{2}-H(Z)H''(Z)\left(Y^{2}-H(Z)^{2}\right)^{2}}{H(Z)^{8}}.\tag{\theequation \emph{f}}
\end{gather}
\end{subequations}
We note that the FENE-CR and Oldroyd-B fluids exhibit a second-order fluid behavior at $O(De)$, so that the velocity field remains Newtonian, i.e., $U_{z,1}=U_{y,1}=0$, following the theorem of Tanner and Pipkin \citep{tanner1966plane,tanner1969intrinsic}. 

\subsection{Second-order solution for the pressure drop of a FENE-CR fluid}

At the second order, $O(De^2)$, the governing equations (\ref{Non-dim Gov.Eq FCR}) yield
\begin{subequations}
\begin{equation}
    \frac{\partial U_{z,2}}{\partial Z}+\frac{\partial U_{y,2}}{\partial Y}=0, \label{Second-order New-Gov.Eq FCR a}
\end{equation} 
\begin{equation}
    \frac{\mathrm{d} P_2}{\mathrm{d} Z}=(1-\tilde\beta)\frac{\partial^2 U_{z,2}}{\partial Y^2}+\tilde\beta\left[ \frac{\partial \tilde A_{zz,3}}{\partial Z}+\frac{\partial}{\partial Y}\left(\tilde A_{yz,3}+\frac{\tilde A_{yz,1} \tilde A_{zz,2}}{L^2 \epsilon ^2}\right)\right],  \label{Second-order New-Gov.Eq FCR b}
\end{equation}
\begin{equation}
U_{z,0}\frac{\partial \tilde A_{zz,2}}{\partial Z}+U_{y,0}\frac{\partial \tilde A_{zz,2}}{\partial Y}-2 \frac{\partial U_{z,0}}{\partial Z}\tilde A_{zz,2}-2 \frac{\partial U_{z,0}}{\partial Y}\tilde A_{yz,2}=-\tilde A_{zz,3},  \label{Second-order New-Gov.Eq FCR c}
\end{equation}
\begin{equation}
U_{z,0}\frac{\partial \tilde A_{yz,2}}{\partial Z}+U_{y,0}\frac{\partial \tilde A_{yz,2}}{\partial Y}-\frac{\partial U_{y,0}}{\partial Z}\tilde A_{zz,2}-\frac{\partial U_{z,0}}{\partial Y}\tilde A_{yy,2}-\frac{\partial U_{z,2}}{\partial Y}=-\tilde A_{yz,3}-\frac{\tilde A_{yz,1}\tilde A_{zz,2}}{L^2 \epsilon ^2}, \label{Second-order New-Gov.Eq FCR d}
\end{equation}
\begin{equation}
U_{z,0}\frac{\partial \tilde A_{yy,2}}{\partial Z}+U_{y,0}\frac{\partial \tilde A_{yy,2}}{\partial Y}-2\frac{\partial U_{y,0}}{\partial Z}\tilde A_{yz,2}-2\frac{\partial U_{ y,0}}{\partial Y}\tilde A_{yy,2}-2\frac{\partial U_{y,2}}{\partial Y}=-\tilde A_{yy,3}-\frac{\tilde A_{yy,1}\tilde A_{zz,2}}{L^2 \epsilon ^2}, \label{Second-order New-Gov.Eq FCR e}
\end{equation}\label{Second-order New-Gov.Eq FCR}\end{subequations}
where we have used the expressions $\tilde A_{zz,1}=0$, $U_{z,1}=0$, and $U_{y,1}=0$. The governing equations (\ref{Second-order New-Gov.Eq FCR}) are subject to the boundary conditions 
\refstepcounter{equation}
$$
 U_{z,2}(H(Z),Z)=0, \; U_{y,2}(H(Z),Z)=0, \; \frac{\partial U_{z,2}}{\partial Y}(0,Z)=0, \; \int^{H(Z)}_0 U_{z,2}(Y,Z)\mathrm{d}Y=0. \eqno{(\theequation{a\mbox{$-$}d})}\label{Second-order BC FCR}
$$
We note that the evolution equation for $\tilde A_{zz,3}$, given in (\ref{Second-order New-Gov.Eq FCR c}), is the same for the FENE-CR and Oldroyd-B fluids. In contrast, the evolution equations for $\tilde A_{yz,3}$ and $\tilde A_{yy,3}$, given in (\ref{Second-order New-Gov.Eq FCR d}) and (\ref{Second-order New-Gov.Eq FCR e}), are different for the two fluids due to additional terms for the FENE-CR fluid, which depend on $L^2 \epsilon^2$. 
Nevertheless, similar to the first order, the expressions for the velocity and pressure drop at $O(De^2)$ are the same for the FENE-CR and Oldroyd-B fluids. This can be seen by substituting (\ref{Second-order New-Gov.Eq FCR d}) into the last term on the right-hand side of the momentum equation~(\ref{Second-order New-Gov.Eq FCR b}), thus clearly showing that the velocity and pressure are independent of $L^2 \epsilon^2$ at $O(De^2)$.

The resulting expressions for $U_{z,2}$, $U_{y,2}$ and
$\tilde A_{zz,3}$, $\tilde A_{yz,3}$, $\tilde A_{yy,3}$ are readily found using \textsc{Mathematica}, but they are rather lengthy and, thus, not presented here.
As the $\tilde A_{yz,3}$ and $\tilde A_{yy,3}$ for the FENE-CR fluid are coupled to $L^2 \epsilon^2$, we expect the pressure drop to depend on the finite extensibility at the next order, $O(De^3)$. We show this dependence in the following subsection.

\subsection{Third-order solution  for the pressure drop of a FENE-CR fluid}

Substituting (\ref{perturbation_expansion FCR}) and (\ref{Series expansion FCR}) into (\ref{Non-dim Gov.Eq FCR}) and considering the third order in $De$, we obtain 
\begin{subequations}
\begin{equation}
    \frac{\partial U_{z,3}}{\partial Z}+\frac{\partial U_{y,3}}{\partial Y}=0, \label{Third-order New-Gov.Eq FCR a}  
\end{equation}
\begin{eqnarray}
\mbox{\hspace{-2mm}} \frac{\mathrm{d} P_3}{\mathrm{d}  Z}&=&(1-\tilde\beta)\frac{\partial^2 U_{z,3}}{\partial Y^2}\nonumber\\
\mbox{\hspace{-2mm}} &+& \tilde\beta\left[ \frac{\partial}{\partial Z}\left(\tilde A_{zz,4}+\frac{(\tilde A_{zz,2})^2}{L^2 \epsilon ^2}\right)+\frac{\partial}{\partial Y}\left(\tilde A_{yz,4}+\frac{\tilde A_{yz,2}\tilde A_{zz,2}+\tilde A_{yz,1}\tilde A_{zz,3}}{L^2 \epsilon ^2}\right)\right],\label{Third-order New-Gov.Eq FCR b} 
\end{eqnarray}
\begin{eqnarray}
U_{z,0}\frac{\partial \tilde A_{zz,3}}{\partial Z}&+&U_{y,0}\frac{\partial \tilde A_{zz,3}}{\partial Y}-2 \frac{\partial U_{z,0}}{\partial Z}\tilde A_{zz,3}-2 \frac{\partial U_{z,0}}{\partial Y}\tilde A_{yz,3} \nonumber\\
&-&2 \frac{\partial U_{z,2}}{\partial Y}\tilde A_{yz,1}=-\left(\tilde A_{zz,4}+\frac{(\tilde A_{zz,2})^2}{L^2 \epsilon ^2}\right), \label{Third-order New-Gov.Eq FCR c} 
\end{eqnarray}
\begin{eqnarray}
U_{z,0}\frac{\partial \tilde A_{yz,3}}{\partial Z}&+&U_{z,2}\frac{\partial \tilde A_{yz,1}}{\partial Z}+U_{y,0}\frac{\partial \tilde A_{yz,3}}{\partial Y}+U_{y,2}\frac{\partial \tilde A_{yz,1}}{\partial Y}-\frac{\partial U_{y,0}}{\partial Z}\tilde A_{zz,3}-\frac{\partial U_{z,0}}{\partial Y}\tilde A_{yy,3}  \nonumber\\
&-&\frac{\partial U_{z,2}}{\partial Y}\tilde A_{yy,1}-\frac{\partial U_{z,3}}{\partial Y}=-\left(\tilde A_{yz,4}+\frac{\tilde A_{yz,2}\tilde A_{zz,2}+\tilde A_{yz,1}\tilde A_{zz,3}}{L^2 \epsilon ^2}\right), \label{Third-order New-Gov.Eq FCR d} 
\end{eqnarray}
\begin{eqnarray}
&&U_{z,0}\frac{\partial \tilde A_{yy,3}}{\partial Z}+U_{z,2}\frac{\partial \tilde A_{yy,1}}{\partial Z}+U_{y,0}\frac{\partial \tilde A_{yy,3}}{\partial Y}+U_{y,2}\frac{\partial \tilde A_{yy,1}}{\partial Y}-2\frac{\partial U_{y,0}}{\partial Z}\tilde A_{yz,3}-2\frac{\partial U_{y,2}}{\partial Z}\tilde A_{yz,1}\nonumber\\
&&\mbox{\hspace{-4mm}}-2\frac{\partial U_{y,0}}{\partial Y}\tilde A_{yy,3}-2\frac{\partial U_{y,2}}{\partial Y}\tilde A_{yy,1}-2\frac{\partial U_{y,3}}{\partial Y}=-\left(\tilde A_{yy,4}+\frac{\tilde A_{yy,2}\tilde A_{zz,2}+\tilde A_{yy,1}\tilde A_{zz,3}}{L^2 \epsilon ^2}\right), \label{Third-order New-Gov.Eq FCR e} 
\end{eqnarray}\label{Third-order New-Gov.Eq FCR}\end{subequations}
where we have used the expressions $\tilde A_{zz,1}=0$, $U_{z,1}=0$ and $U_{y,1}=0$. The governing equations (\ref{Third-order New-Gov.Eq FCR}) are subject to the boundary conditions 
\refstepcounter{equation}
$$
 U_{z,3}(H(Z),Z)=0, \; U_{y,3}(H(Z),Z)=0, \; \frac{\partial U_{z,3}}{\partial Y}(0,Z)=0, \; \int^{H(Z)}_0 U_{z,3}(Y,Z)\mathrm{d}Y=0. \eqno{(\theequation{a\mbox{$-$}d})}\label{Third-order BC FCR}
$$
First, we integrate (\ref{Third-order New-Gov.Eq FCR b}) twice with respect to $Y$ and apply the boundary conditions (\ref{Third-order BC FCR}$a$) and (\ref{Third-order BC FCR}$c$), to obtain the expression for $U_{z,3}(Y,Z)$ that involves the pressure gradient $\mathrm{d}P_3/\mathrm{d}Z$. The resulting expression is lengthy and thus not shown here. 
To determine $\mathrm{d}P_3/\mathrm{d}Z$, we use the integral constraint (\ref{Third-order BC FCR}$d$), leading to 
\begin{eqnarray}
 \mbox{\hspace{-2mm}} \frac{\mathrm{d}P_3}{\mathrm{d}Z}&=&\frac{10692 \tilde\beta H'(Z)}{35 L^2 \epsilon ^2 H(Z)^9} \nonumber\\
&&\mbox{\hspace{-8mm}}+\frac{216 \tilde\beta}{35}\left[(\tilde\beta-8)\frac{H'''(Z)}{H(Z)^7}+ (110-13 \tilde\beta) \frac{H'(Z) H''(Z)}{H(Z)^8}+24 (\tilde\beta-9) \frac{H'(Z)^3}{H(Z)^9}\right].
\label{Third-order Pgrad FCR}
\end{eqnarray}
Integrating (\ref{Third-order Pgrad FCR}) with respect to $Z$ from 0 to 1 provides an expression for the pressure drop of the FENE-CR fluid at $O(De^3)$ given in (\ref{pressure drop sol third order FCR}).

\subsection{Fourth-order solution for the pressure drop of a FENE-CR fluid}

To calculate the pressure drop at the next order, $O(De^4)$, we use the expression (\ref{dP non-uniform by parts}), which resembles the result of an application of the reciprocal theorem~\citep{boyko2021RT,boyko2022pressure}, and requires only the knowledge of velocity and conformation tensor components from the previous orders. At $O(De^4)$, the expression for the pressure drop $\Delta P_4$ takes the form 
\begin{eqnarray}
\Delta P_4&=&\tilde\beta \int^{H(0)}_0[\mathcal{G}_{zz,5}\hat U_z]_{Z=0} \mathrm{d}Y-\tilde\beta \int^{H(1)}_0[\mathcal{G}_{zz,5}\hat U_z]_{Z=1} \mathrm{d}Y \nonumber\\
&&+\tilde\beta \int^1_0\int^{H(Z)}_0\left(\mathcal{G}_{zz,5}\frac{\partial \hat U_z}{\partial Z}+\mathcal{G}_{yz,5}\frac{\partial \hat U_z}{\partial Y}\right) \mathrm{d}Y \mathrm{d}Z,
\label{Pdrop_reciprocal_FCR_4thO}
\end{eqnarray}
where $\mathcal{G}_{zz,5}$ and $\mathcal{G}_{yz,5}$ are given by
\begin{subequations}
    \begin{eqnarray}\label{Gzz5 FCR}
    \mathcal{G}_{zz,5}&=&-U_{z,0}\frac{\partial \tilde A_{zz,4}}{\partial Z}-U_{z,2}\frac{\partial \tilde A_{zz,2}}{\partial Z}-U_{y,0}\frac{\partial \tilde A_{zz,4}}{\partial Y}-U_{y,2}\frac{\partial \tilde A_{zz,2}}{\partial Y}+2 \frac{\partial U_{z,0}}{\partial Z}\tilde A_{zz,4} \nonumber \\
    &&+2 \frac{\partial U_{z,2}}{\partial Z}\tilde A_{zz,2}+2 \frac{\partial U_{z,0}}{\partial Y}\tilde A_{yz,4}+2 \frac{\partial U_{z,2}}{\partial Y}\tilde A_{yz,2}+2 \frac{\partial U_{z,3}}{\partial Y}\tilde A_{yz,1},
\end{eqnarray}
\begin{eqnarray}\label{Gyz5 FCR}
    \mathcal{G}_{yz,5}&=& -U_{z,0}\frac{\partial \tilde A_{yz,4}}{\partial Z}-U_{z,2}\frac{\partial \tilde A_{yz,2}}{\partial Z}-U_{z,3}\frac{\partial \tilde A_{yz,1}}{\partial Z}-U_{y,0}\frac{\partial \tilde A_{yz,4}}{\partial Y}-U_{y,2}\frac{\partial \tilde A_{yz,2}}{\partial Y}-U_{y,3}\frac{\partial \tilde A_{yz,1}}{\partial Y} \nonumber \\
    &+&\frac{\partial U_{y,0}}{\partial Z}\tilde A_{zz,4}+\frac{\partial U_{y,2}}{\partial Z}\tilde A_{zz,2}+\frac{\partial U_{z,0}}{\partial Y}\tilde A_{yy,4}+\frac{\partial U_{z,2}}{\partial Y}\tilde A_{yy,2}+\frac{\partial U_{z,3}}{\partial Y}\tilde A_{yy,1}+\frac{\partial U_{z,4}}{\partial Y}.
\end{eqnarray}
\end{subequations}
We note that, because of the integral constraint $\int_{0}^{H(Z)}U_{z,4}\mathrm{d}Y=0$, the last term appearing in (\ref{Gyz5 FCR}), $\partial U_{z,4}/\partial Y$, satisfies  
 \begin{equation}\label{higher-order integral constraint}
    \int^{H(Z)}_0\frac{\partial U_{z,4}}{\partial Y}\frac{\partial \hat U_z}{\partial Y} \mathrm{d}Y =-\int^{H(Z)}_0 U_{z,4}\frac{\partial^2 \hat U}{\partial Y^2} \mathrm{d}Y =-\frac{\mathrm{d}\hat P}{\mathrm{d} Z}\int^{H(Z)}_0U_{z,4} \mathrm{d}Y=0,
 \end{equation}
and thus, this term does not contribute to the pressure drop, since it is identically zero. Therefore, the expressions for $\mathcal{G}_{zz,5}$ and $\mathcal{G}_{yz,5}$ depend on the solution from the previous orders, and we can calculate the fourth-order pressure drop $\Delta P_4$ using the results of the leading-, first-, second-, and third-order viscoelastic problems. The resulting expression for $\Delta P_4$ for the FENE-CR fluid is given in (\ref{pressure drop sol fourth order FCR}).

For completeness, in the supplementary material, we provide the \textsc{Mathematica} file containing the explicit expressions for the velocity, conformation tensor components, and the pressure drop in the low-Deborah-number limit up to $O(De^4)$.

\section{Details of numerical simulations using OpenFOAM}\label{Details of numerical simulations using OpenFOAM}

In this appendix, we describe the numerical procedure used to solve the system of nonlinear governing equations (\ref{Continuity+Momentum FCR})--(\ref{Conformation tensor FCR}) for the viscoelastic fluid flow. Besides the FENE-CR fluid, we also consider the Oldroyd-B fluid for comparison and validation.
We have performed two-dimensional finite-volume simulations using an open-source framework OpenFOAM~\citep{jasak2007openfoam} integrated with viscoelastic flow solver \textsc{RheoTool}~\citep{pimenta2017stabilization}. We use the log-conformation method to calculate the polymer stress tensor by solving the equations for the logarithm of the conformation tensor $\boldsymbol{\Theta}$ instead of $\boldsymbol \tau_p$~\citep{pimenta2017stabilization,habla2014numerical,kumar2021numerical,kumar2021elastic}. Under the log-conformation approach, the conformation tensor is positive definite at high Deborah/Weissenberg numbers, ensuring the stability of the numerical solution~\citep{fattal2004constitutive,fattal2005time}. The details of the numerical implementation and the code validation are given in prior studies~\citep[see, e.g.,][]{pimenta2017stabilization,favero2010viscoelastic}. 

In our simulations, we impose the no-slip and no-penetration boundary conditions along the wall, $y= \pm h(z)$, and a fully developed unidirectional Poiseulille velocity profile at the entrance and exit. In addition, we specify a null value of polymeric stress tensor and zero-gradient of pressure at the channel entrance. At the channel wall, we impose a linear extrapolation for polymer stresses and zero-normal gradient for pressure~\citep{pimenta2017stabilization}. At the exit, we use a zero-gradient boundary condition for polymer stresses and prescribe a constant value for pressure, $p=0$. Finally, we calculate the pressure drop along the centreline between the inlet ($z=0$) and outlet ($z=\ell$) of the contraction, i.e. $\Delta p=p(y=0,z=0)-p(y=0,z=\ell)$, eliminating the entrance and exit effects.

\begin{table}
  \begin{center}
\def~{\hphantom{0}}
  \begin{tabular}{ccccccccccc}
   $\ell$ &   $h_{0}$ & $h_{\ell}$ & $\mu_0$ & $\rho$  & $q$  & $u_c$ & $p_c$ & $\lambda$ & $De$ & $\tilde{\beta}$ \\
     (mm)  &  (mm) & (mm) & (Pa s) & (kg m$^{-3}$) & (mm${^2}$ s$^{-1}$)& (mm s$^{-1}$) & (Pa) & (s) & (--)  & (--)\\
     5 & 0.4 & 0.1 & 1 & 1 & 1 & 5 & 2500 & $0-4$ & $0-4$ & 0.05, 0.4\\
    \end{tabular}
  \caption{Values of physical and geometrical parameters used in the two-dimensional numerical simulations of the pressure-driven flow of the FENE-CR fluid in a hyperbolic contracting channel.}
  \label{T2}
  \end{center}
\end{table}
We summarize in table~\ref{T2} the values of physical and geometrical parameters used in the numerical simulations. We consider a geometry with an inlet-to-outlet ratio $H_0=h_0/h_\ell=4$ and an aspect ratio $\epsilon=h_\ell/\ell=0.02$ and explore two different polymer-to-total viscosity ratios: $\tilde{\beta}=\mu_p/\mu_0=0.4$ and $\tilde{\beta}=\mu_p/\mu_0=0.05$, where the latter corresponds to the ultra-dilute limit. 
In all simulations, we keep $\ell_0=\ell$ and $\ell_\ell=5\ell$. 

To study the effect of Deborah numbers in the case of the FENE-CR fluid, we mainly set the finite extensibility parameter to $L^2=1250$, corresponding to $L^2\epsilon^2=0.5$, and change the value of relaxation time $\lambda$, while keeping the values of all other parameters. When investigating the effect of finite extensibility $L^2\epsilon^2$ on the pressure drop, we change the value of $L^2$ and set different $\lambda$ corresponding to different values of $De$, while keeping the values of all other parameters. 
Similarly, when analyzing the pressure drop at different viscosity ratios $\mu_p/\mu_0$, we change the value of $\mu_p$ and $\mu_s$, while setting $\mu_0=1$ Pa s and keeping the values of all other parameters. 
We note that the effect of fluid inertia is negligible in our simulations because the reduced Reynolds number $\epsilon Re=(h_{\ell}/\ell)\rho u_{c}h_{\ell}/\mu_{0}= 10^{-8}$ is very small. Eventually, we use the transient rheoFoam solver~\citep{pimenta2017stabilization} for simulations, and once the residuals of the variables $\boldsymbol{u}$, $p$ and $\boldsymbol{\Theta}$ becomes less than $10^{-6}$, we terminate the simulation and calculate the pressure drop. We non-dimensionalize the time $t$ using the residence time in the contraction $t_c=\ell/u_c=1$ s.
Typical non-dimensional values of the time step are $\Delta T= 10^{-4}$ for the low-$De$ simulations and a reduced time step $\Delta T= 10^{-5}$ for the high-$De$ simulations. 

To assess the grid sensitivity, we have performed tests by considering three different mesh resolutions (total number of node points is 75672, 114882, and 139482) at four different Deborah numbers ($De$ = 1, 2, 3, and 4), and established grid independence with a maximum relative error of 0.3 \% for the pressure drop. We have also carried out numerical simulations without the log-conformation approach and found an excellent agreement with the log-conformation results.

In addition, we cross-validate our OpenFOAM results for Oldroyd-B and FENE-CR fluids with those obtained from the finite-element software COMSOL Multiphysics. The details of the numerical implementation in COMSOL are given by~\citet{boyko2022pressure} for the Oldroyd-B fluid. To simulate the FENE-CR fluid in COMSOL, we impose the polymer stress distribution corresponding to the Poiseuille flow at the entrance. 

\begin{figure}
 \centerline{\includegraphics[scale=1]{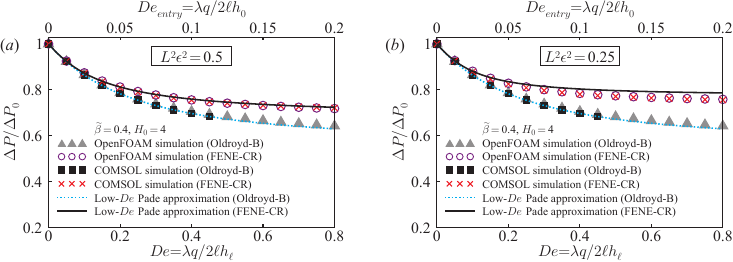}}
\caption{Comparison of simulation results obtained from OpenFOAM and COMSOL for the pressure drop for the Oldroyd-B and FENE-CR fluids in a contracting channel. ($a,b$) Scaled pressure drop $\Delta P/\Delta P_0$ as a function of $De=\lambda q/(2\ell h_{\ell})$ for ($a$) $L^2\epsilon^2=0.5$ and ($b$) $L^2\epsilon^2=0.25$. Gray triangles and purple circles represent the OpenFOAM simulation results for the Oldroyd-B and FENE-CR fluids. Black squares and red crosses represent the COMSOL simulation results for the Oldroyd-B and FENE-CR fluids. Cyan dotted and solid black lines represent the low-$De$ Pad\'{e} approximation (\ref{Pade approximation dP}) applied to the fourth-order asymptotic solutions for the Oldroyd-B and FENE-CR fluids. All calculations were performed using $H_0=4$ and $\tilde\beta=0.4$.}\label{Fig:8}
\end{figure}

We present in figure~\ref{Fig:8} the scaled pressure drop $\Delta P/\Delta P_0$ for the Oldroyd-B and FENE-CR fluids as a function of $De$ in a contracting channel for $L^2\epsilon^2=0.5$ ($a$) and $L^2\epsilon^2=0.25$ ($b$), with $H_0=4$ and $\tilde\beta=0.4$. Gray triangles and purple circles represent the OpenFOAM simulation results for the Oldroyd-B and FENE-CR fluids. Black squares and red crosses represent the COMSOL simulation results for the Oldroyd-B and FENE-CR fluids. Cyan dotted and solid black lines represent the low-$De$ Pad\'{e} approximation (\ref{Pade approximation dP}) for the Oldroyd-B and FENE-CR fluids. 
In COMSOL simulations of the Oldroyd-B fluid, we could not obtain converged results beyond $De=0.45$. In contrast, using OpenFOAM, we have performed simulations up to $De=4$ with no difficulties, thus achieving the high-$De$ limit. We encountered no convergence issues when running simulations with the FENE-CR model in OpenFOAM (up to $De=4$) and COMSOL (up to $De=0.8$).
Clearly, for both Oldroyd-B and FENE-CR fluids, there is excellent agreement between the simulation results obtained from OpenFOAM and COMSOL. 
In particular, for the Oldroyd-B fluid, the maximum relative error is 1.3 \% at $De=0.45$. Similarly, for the FENE-CR fluid, we find a maximum relative error of 0.4 \% and 0.3 \% at $De=0.8$, corresponding to $L^2\epsilon^2=0.5$ and $L^2\epsilon^2=0.25$, respectively.

\bibliographystyle{jfm}
\bibliography{literature}

\end{document}